\documentclass{aa}

\newcommand{\mm}{\, \mu \mathrm{m}}

\newcommand{\kms}{\, \mathrm{km}\,\mathrm{s}^{-1}}
\newcommand{\Mpc}{\, \mathrm{Mpc}}

\newcommand{\new}[1]{{#1}}
\newcommand{\nnew}[1]{{#1}}

\usepackage{url}
\usepackage{graphicx}
\usepackage[varg]{txfonts}
\usepackage{amsmath}
\usepackage{amssymb}
\usepackage{wasysym}
\usepackage{pdfpages}
\usepackage{natbib}
\usepackage{multirow}
\usepackage{booktabs}
\usepackage{color}
\usepackage[caption=false]{subfig}

\begin{document}

\title{A low-luminosity type-1 QSO sample} 
\subtitle{V. Overluminous host spheroids and their excitation mechanisms\thanks{Based on observations with ESO-NTT, proposal no. 83.B-0739}}

\author
	{
	Gerold Busch\inst{1},
	Nastaran Fazeli\inst{1},
	Andreas Eckart\inst{1,2},
	M\'onica Valencia-S.\inst{1},
	Semir Smaji\'c\inst{1,2},
	Lydia Moser\inst{2,1},
	Julia Scharw\"achter\inst{3},
	Jens Dierkes\inst{4},
	Sebastian Fischer\inst{5}
	}
	
\institute{
I. Physikalisches Institut der Universit\"at zu K\"oln, Z\"ulpicher Str. 77, 50937 K\"oln, Germany\\
\email{busch@ph1.uni-koeln.de}
\and
Max-Planck-Institut f\"ur Radioastronomie, Auf dem H\"ugel 69, 53121 Bonn, Germany
\and
LERMA, Observatoire de Paris, PSL, CNRS, Sorbonne Universit\'es, UPMC, F-75014, Paris, France
\and
Nieders\"achsische Staats- und Universit\"atsbibliothek G\"ottingen, 37070 G\"ottingen, Germany
\and
Deutsches Zentrum f\"ur Luft- und Raumfahrt (DLR), K\"onigswinterer Str. 522-524, 53227 Bonn, Germany
}

\date{Accepted ???. Received ???; in original form ???}

\abstract{
We present near-infrared (NIR) $H+K$-band longslit spectra of eleven galaxies which are obtained with SOFI at the NTT (ESO). The galaxies are chosen from the \emph{low-luminosity type-1 quasi-stellar object (LLQSO) sample} which comprises the 99 closest ($z\leq 0.06$) QSOs from the Hamburg/ESO survey for bright UV-excess QSOs. These objects are ideal targets to study the gap between local Seyfert galaxies and high-redshift quasars, since they show much stronger AGN activity compared to local objects but are still close enough for a detailed structural analysis. 

We fit hydrogen recombination, molecular hydrogen, and [\ion{Fe}{ii}] lines after carefully subtracting the continuum emission. From the broad Pa$\alpha$ components, we estimate black hole masses and enlarge the sample of LLQSOs that show a deviation from the $M_\mathrm{BH}-L_\mathrm{bulge}$ relations of inactive galaxies from 12 to 16 objects.

All objects show emission from hot dust ($T\sim 1200\,\mathrm{K}$) as well as stellar contribution. However, the particular fractions vary a lot between the objects. More than half of the objects show H$_2$ emission lines that are indicating a large reservoir of molecular gas which is needed to feed the AGN and star formation. 
\nnew{
In the NIR diagnostic diagram all objects lie in the location of AGN dominated objects. However, most of the objects show indications of star formation activity, suggesting that their offset location with respect to $M_\mathrm{BH}-L_\mathrm{bulge}$ relations of inactive galaxies may be a consequence of overluminous bulges.
}
}


\titlerunning{LLQSO sample: V. NIR spectroscopy}
\authorrunning{Gerold Busch et al.}

\maketitle

\section{Introduction}

Numerous studies have shown that the mass of the supermassive black hole (BH), which is believed to be hosted in the center of every galaxy, correlates well with several properties of the host galaxy or at least its central spheroidal component \citep[e.g.,][]{1995ARA&A..33..581K,1998AJ....115.2285M,2000ApJ...539L..13G,2000ApJ...539L...9F,2003ApJ...589L..21M,2004ApJ...604L..89H,2007ApJ...655...77G,2012ApJ...746..113G,2013ARA&A..51..511K,2013MNRAS.434..387S,2014ApJ...780...70L}.
The role of active galactic nuclei (AGN) in this context is still unclear.

\begin{figure}
\centering
\includegraphics[width=\columnwidth]{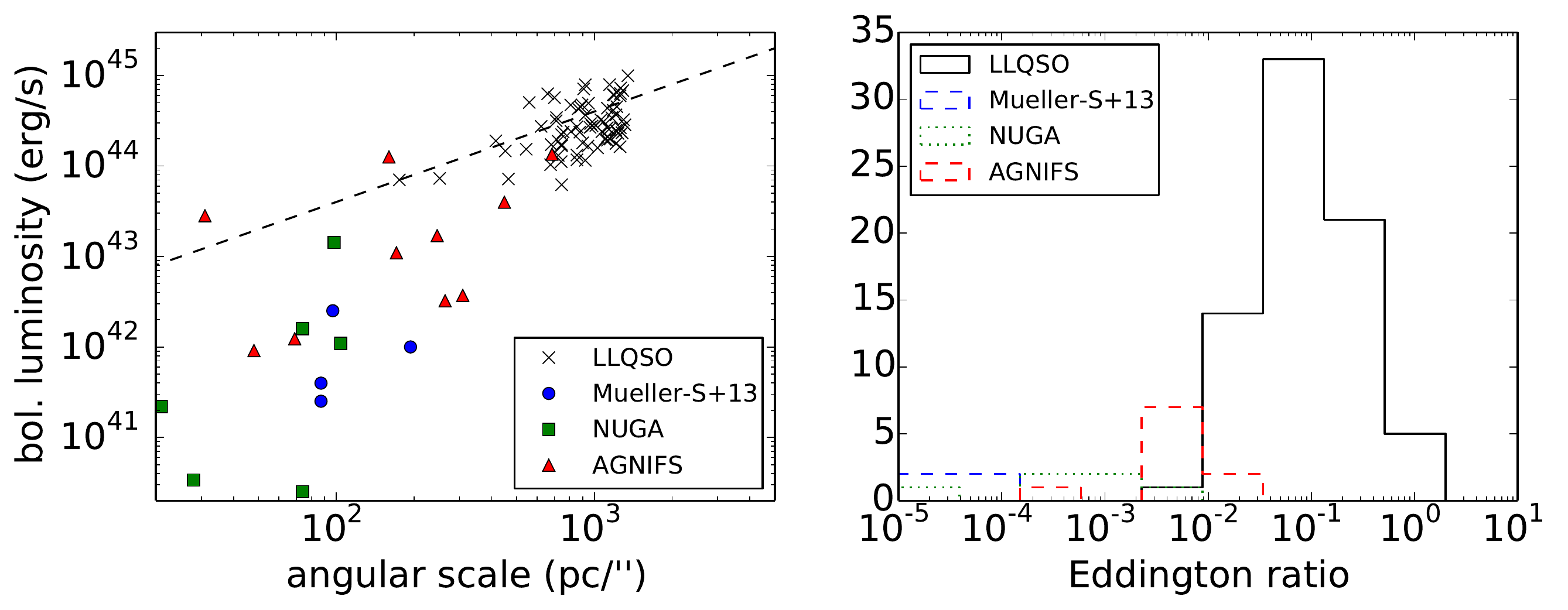}
\caption{\new{\emph{Left:} Bolometric luminosity as a function of the linear scale pc/arcsec. \emph{Right:} Histogram of the Eddington ratios of the LLQSO sample in comparison to the recent study of \cite{2013ApJ...763L...1M}, the NUGA sample \citep{2003A&A...407..485G}, and the work of the AGNIFS group (Storchi-Bergmann, Riffel, et al.).}}
\label{fig:sample}
\end{figure}

\nnew{
AGN feedback has been suggested as a possible regulating mechanism between BHs and their host galaxies, which could contribute to the formation of the local scaling relations through the quenching of star formation. Studies that help to understand the interplay of star formation, black hole accretion and outflows in AGN host galaxies are therefore of high importance.
}
\new{
Over the last years, several studies using integral-field spectroscopy (IFS) in the near-infrared (NIR) have shown that many active galaxies show recent or on-going star formation in the central kiloparsec \citep[e.g.][]{2008AJ....135..479B,2009MNRAS.393..783R,2009ApJ...698.1852B,2012A&A...544A.129V,2014MNRAS.438..329F,2015A&A...575A.128B,2015arXiv150802664S}.
Also, IFS has allowed to spatially resolve inflows and outflows in many galaxies \citep[e.g.][]{2008MNRAS.385.1129R,2010MNRAS.402..819S,2011ApJ...739...69M,2014ApJ...792..101D,2015MNRAS.451.3587R,2015MNRAS.453.1727D}.

From a sample of $\sim$ 10 galaxies, the AGNIFS team (Storchi-Bergmann, Riffel, and collaborators) find that the molecular gas is mostly situated in a disk-like structure and often shows inflow patterns (``feeding'') while the ionized gas is often more perturbed and more affected by outflows from the AGN (``feedback''). 

It is still under debate whether outflows have positive or negative influence on star formation. Most probable is that outflows can be responsible for both: initiating and quenching star formation \citep[e.g.][and references therein]{2010A&A...521A..65N,2012MNRAS.425L..66M,2013ApJ...772..112S,2013A&A...558A...5R,2015ApJ...799...82C,2015A&A...582A..63C}. Furthermore, it has been shown that the presence of a powerful AGN can significantly boost the outflow rate \citep{2014A&A...562A..21C}.

Most of these studies are based on nearby Seyfert galaxies or low-luminosity AGN (redshift $z \lesssim 0.01$). 
\nnew{
However, these objects may not be representative of the higher-redshift AGN population, since the AGN power (and star formation rate) is expected to increase with redshift.
}
With today's instrumentation, it is not possible to resolve the centers of AGNs on sub-kpc scale at the peak of AGN and star formation activity \citep[$z\sim 2$; e.g.,][]{2015MNRAS.451.1892A}.


\nnew{
The \emph{low-luminosity type-1 QSO sample} (LLQSO sample), was selected in order to fill the gap between the local Seyfert population and more powerful QSOs at higher cosmological distances. It is a subsample of the Hamburg/ESO survey \citep[HES;][]{2000A&A...358...77W} and contains only the closest 99 objects with $z\leq 0.06$ which are still close enough to achieve a sub-kpc spatial resolution. 
}

The bolometric luminosity of LLQSOs is systematically higher by at least a magnitude compared to other AGN samples at lower redshift which compensates the lower physical resolution due to the higher redshift. Furthermore, the Eddington ratio which traces the BH accretion rate, is higher by up to several magnitudes (Fig.~\ref{fig:sample}). Therefore, the impact of the AGN on the surrounding interstellar medium (``feedback'') is expected to be much stronger than in local low-luminosity AGN. This makes LLQSOs the ideal targets to study the interplay between the central engine and the host galaxy \citep{2012nsgq.confE..69M}.
}
A more detailed description about the sample can be found in the previous near-infrared imaging study \citep{2014A&A...561A.140B}. Several sources have already been observed in molecular gas \citep{2007A&A...470..571B}, \ion{H}{i} \citep{2009A&A...507..757K}, and H$_2$O-maser emission \citep{2012MNRAS.420.2263K}. A previous NIR spectroscopy study was done by \citet{2006A&A...452..827F}, while single objects from the sample have been studied in \citet{2007A&A...464..187K}, \citet{2011AJ....142...43S}, \citet{2015A&A...575A.128B}, and Moser et al. (in press). A large fraction of the LLQSOs are included in the new \emph{Close AGN Reference Survey} (CARS, B.~Husemann in prep., http://www.cars-survey.org/).

In this paper, we present and analyse NIR spectroscopic observations of eleven sources from the LLQSO sample, 
\nnew{
for which imaging data were discussed in \citet{2014A&A...561A.140B}.
LLQSOs were found to lie below the $M_\mathrm{BH}-L_\mathrm{bulge}$ relations of inactive galaxies, as a possible consequence of enhanced star formation (overluminous bulges) or undermassive BHs
} 
\citep[deviations of active galaxies from the $M_\mathrm{BH}-L_\mathrm{bulge}$ relations have been found in the optical by][]{2004ApJ...615..652N,2008ApJ...687..767K,2011ApJ...726...59B,2012ApJ...757..125U}.
\nnew{
Recent NIR integral-field spectroscopy of the LLQSO HE 1029--1831 shows that, at least in this particular case, the deviation is caused by an overluminosity of the bulge due to young stellar populations \citep{2015A&A...575A.128B}.
}

\new{
Near-infrared spectroscopy is a useful tool to \nnew{assess} extinction, the \nnew{dominating} stellar populations, excitation mechanisms, and the contributions of stellar and non-stellar components to the emission of the galaxy \citep[e.g.][Fazeli et al., in prep.]{2007A&A...466..451Z,2012A&A...544A.105S,2012A&A...544A.129V,2014A&A...567A.119S,2015A&A...575A.128B}.
\nnew{The NIR $H+K$-band contains several diagnostic lines:}
Hydrogen recombination lines (Pa$\alpha$ and Br$\gamma$) can be excited by the AGN (in the broad and narrow line region) but are also tracers of young star formation. Shocks can be traced by molecular hydrogen (H$_2$) rotational-vibrational lines and the forbidden [\ion{Fe}{ii}] line. Stellar CO absorption bands ($^{12}$CO(6-3) in the $H$-band and $^{12}$CO(2-0) in the $K$-band) give constraints on the stellar population, while the forbidden [\ion{Si}{vi}] line is a clear AGN tracer \citep[e.g.][]{riffel_0.8-2.4_2006,mason_nuclear_2015}.

Molecular hydrogen H$_2$ emission is either of thermal or non-thermal origin. The main thermal excitation mechanisms are: shocks, UV radiation in dense clouds, or X-rays \citep[e.g.,][]{1989MNRAS.236..929B,1989ApJ...338..197S,1990ApJ...363..464D} while UV-pumping \citep{1987ApJ...322..412B} is a possible non-thermal excitation mechanism. The H$_2$ line ratio 2-1 S(1)/1-0 S(1) can be used to distinguish between thermal and non-thermal excitation, while the [\ion{Fe}{ii}]/Br$\gamma$ ratio can be used to estimate the importance of X-ray excitation \citep[e.g.][]{1993ApJ...411..565C,1997ApJ...482..747A,2002MNRAS.331..154R,2003MNRAS.343..192R}.

\nnew{A number of studies have focussed on developing} a diagnostic diagram in the NIR (comparable to the BPT-diagram in the optical) that distinguishes between excitation from star formation, AGN, and shocks. For this, line ratios between shock tracers (H$_2$ or [\ion{Fe}{ii}]) and star formation tracers (Pa$\alpha$, Pa$\beta$, Br$\gamma$) are used \citep[e.g.,][]{1998ApJS..114...59L,2004A&A...425..457R,2005MNRAS.364.1041R,2013MNRAS.430.2002R}. \nnew{Using integral-field spectroscopy, \cite{2015A&A...578A..48C} could spatially separate line emitting regions with different ionization mechanisms (AGN/young stars/supernova dominated) and show that they occupy different regions in the $\log([\ion{Fe}{ii}/\mathrm{Br}\gamma) - \log(\mathrm{H}_2/\mathrm{Br}\gamma)$ diagram.} The observed line ratios could be well reproduced with photoionization models by \cite{2012MNRAS.422..252D}. Furthermore, they find that X-ray emission from the AGN can be considered as most important excitation mechanisms of these lines.
}

\nnew{Here, we use $H+K$ longslit spectra to study extinction, star formation activity, and black-hole masses via NIR diagnostic lines in a sample of 11 LLQSOs.}
This paper is structured as follows: In Sect.~\ref{sec:obs}, \nnew{we describe the observations as well as the data reduction and calibration}. In Sect.~\ref{sec:continuum}, we explain our continuum subtraction, followed by a description of the emission line fits which are based on the continuum subtracted spectra from \nnew{Sect.~\ref{sec:continuum}}. Excitation mechanism are discussed in Sect.~\ref{sec:excitation}. In Sect.~\ref{sec:bhmass} we estimate black hole masses and discuss the positions of the galaxies in the black hole mass - bulge luminosity relation. \new{In Sect.~\ref{sec:firprop} we discuss the far-infrared (FIR) properties and finally draw conclusions in Sect.~\ref{sec:conclusions}. Individual objects are discussed in the Appendix.}

\begin{figure}
\centering
\includegraphics[width=0.3\columnwidth]{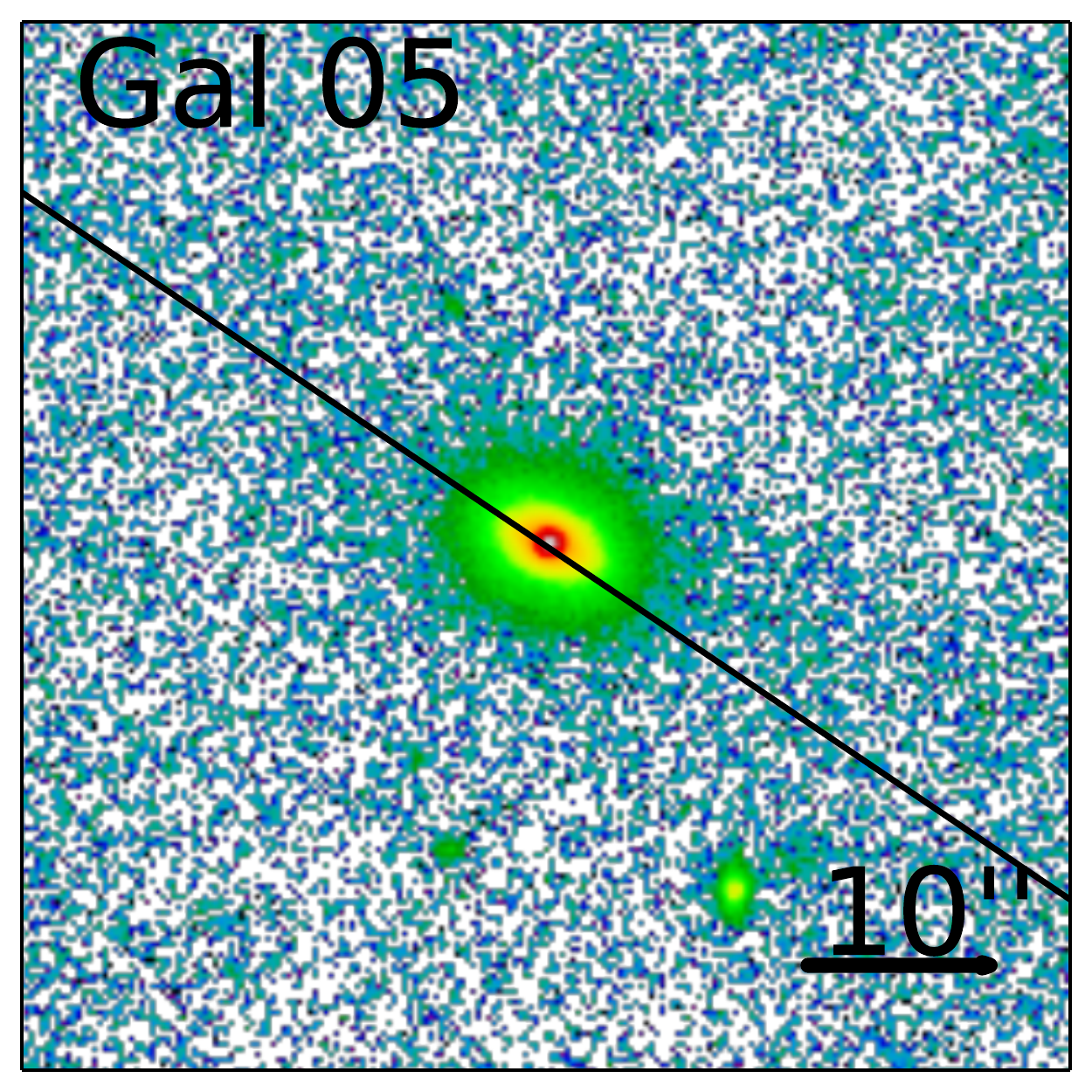}
\includegraphics[width=0.3\columnwidth]{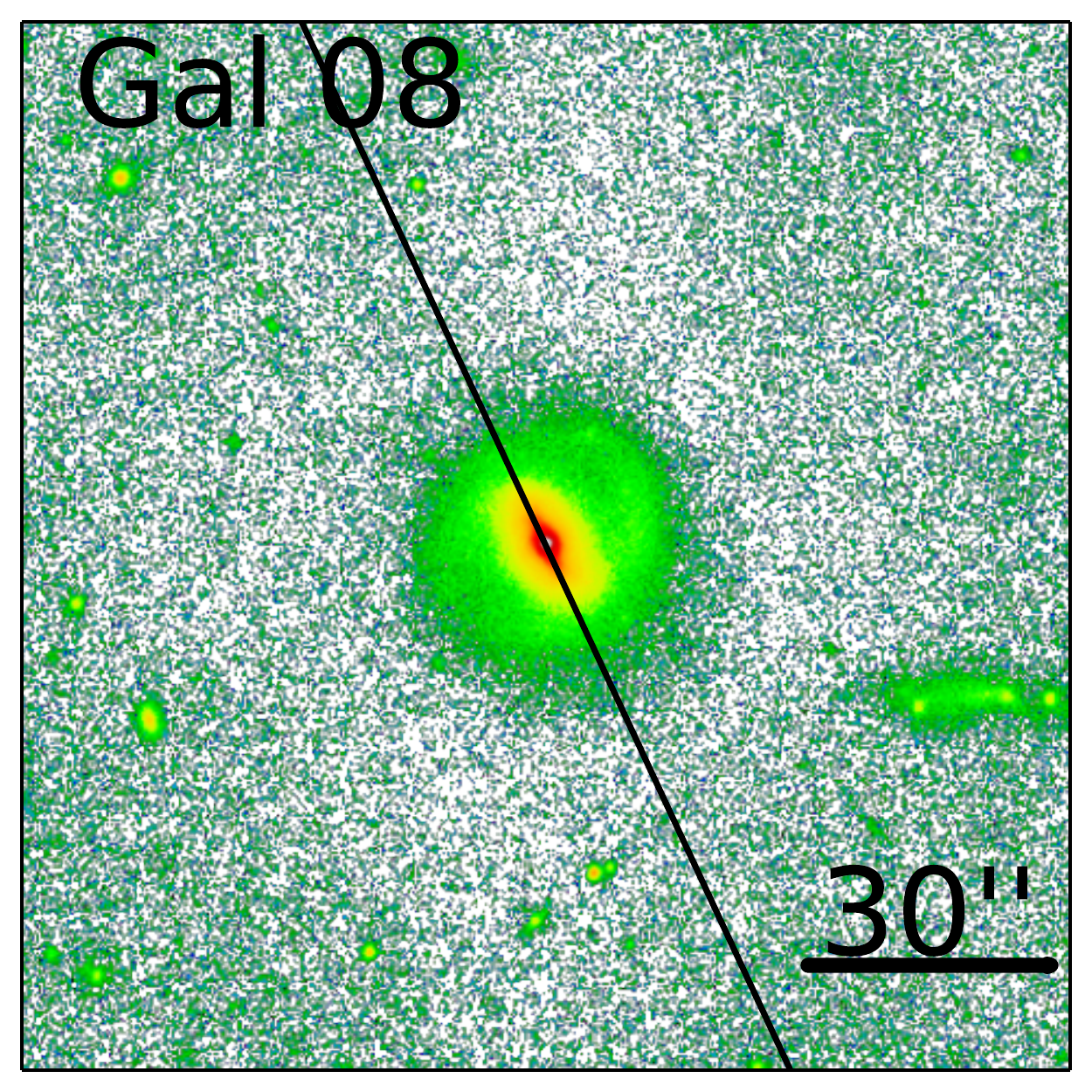}
\includegraphics[width=0.3\columnwidth]{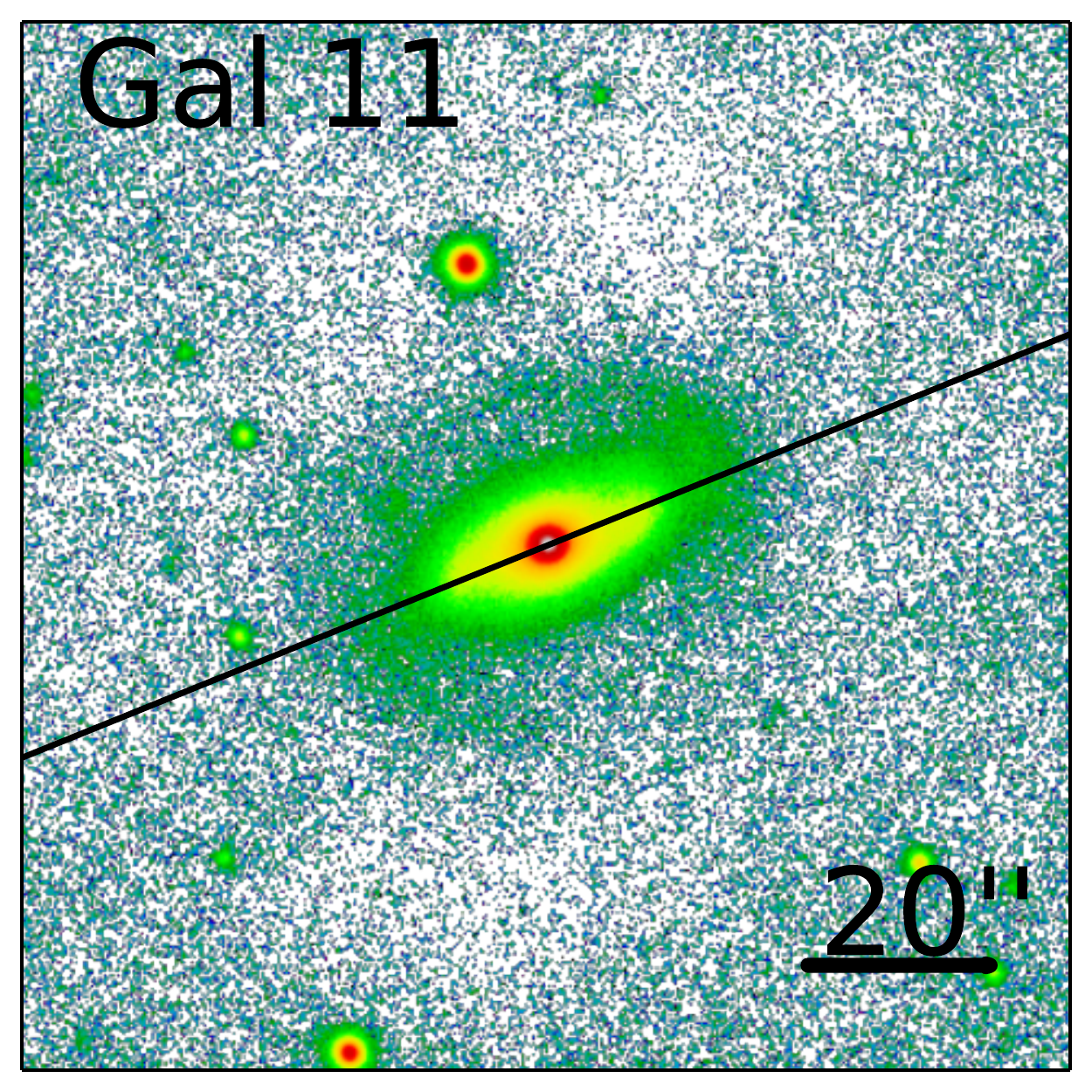}

\includegraphics[width=0.3\columnwidth]{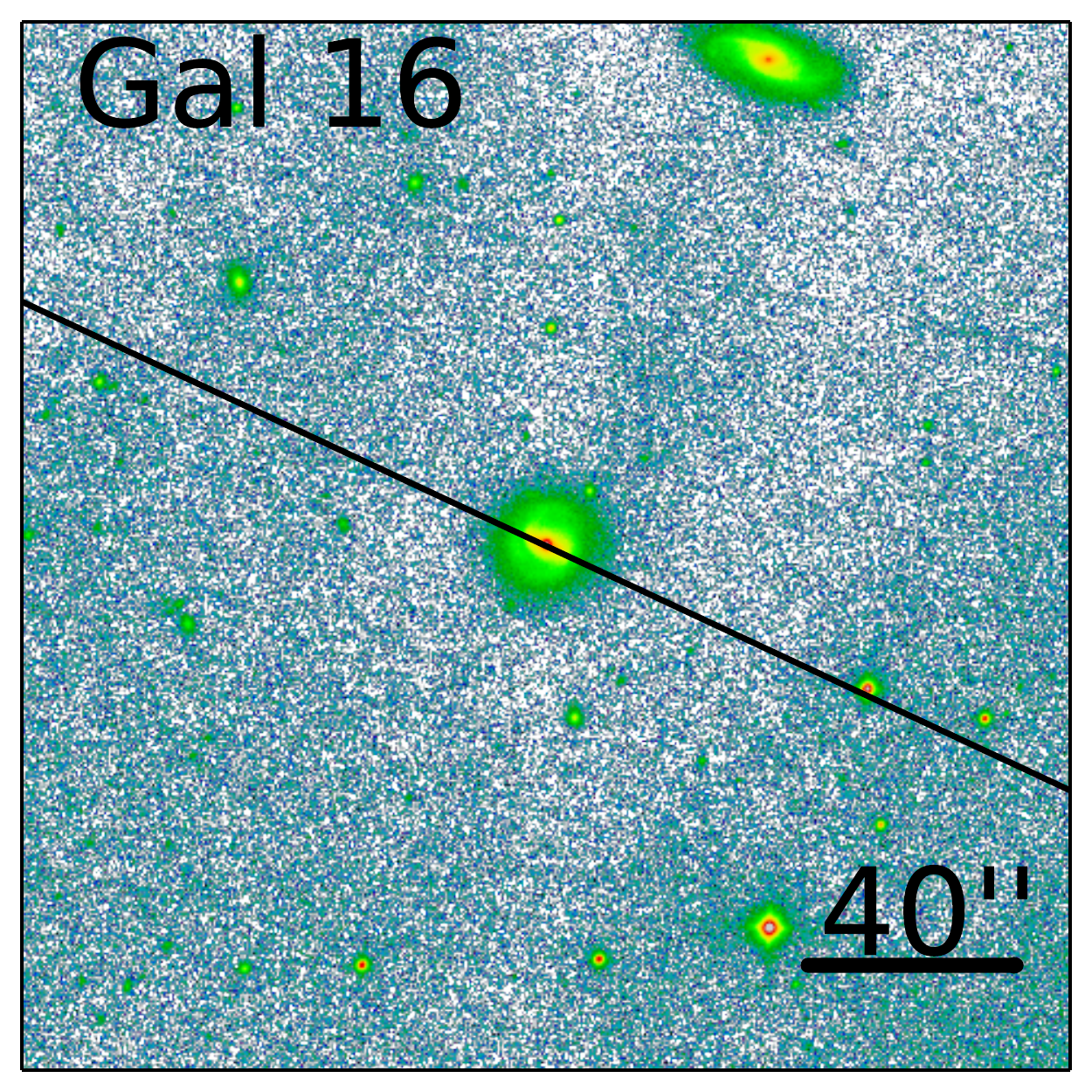}
\includegraphics[width=0.3\columnwidth]{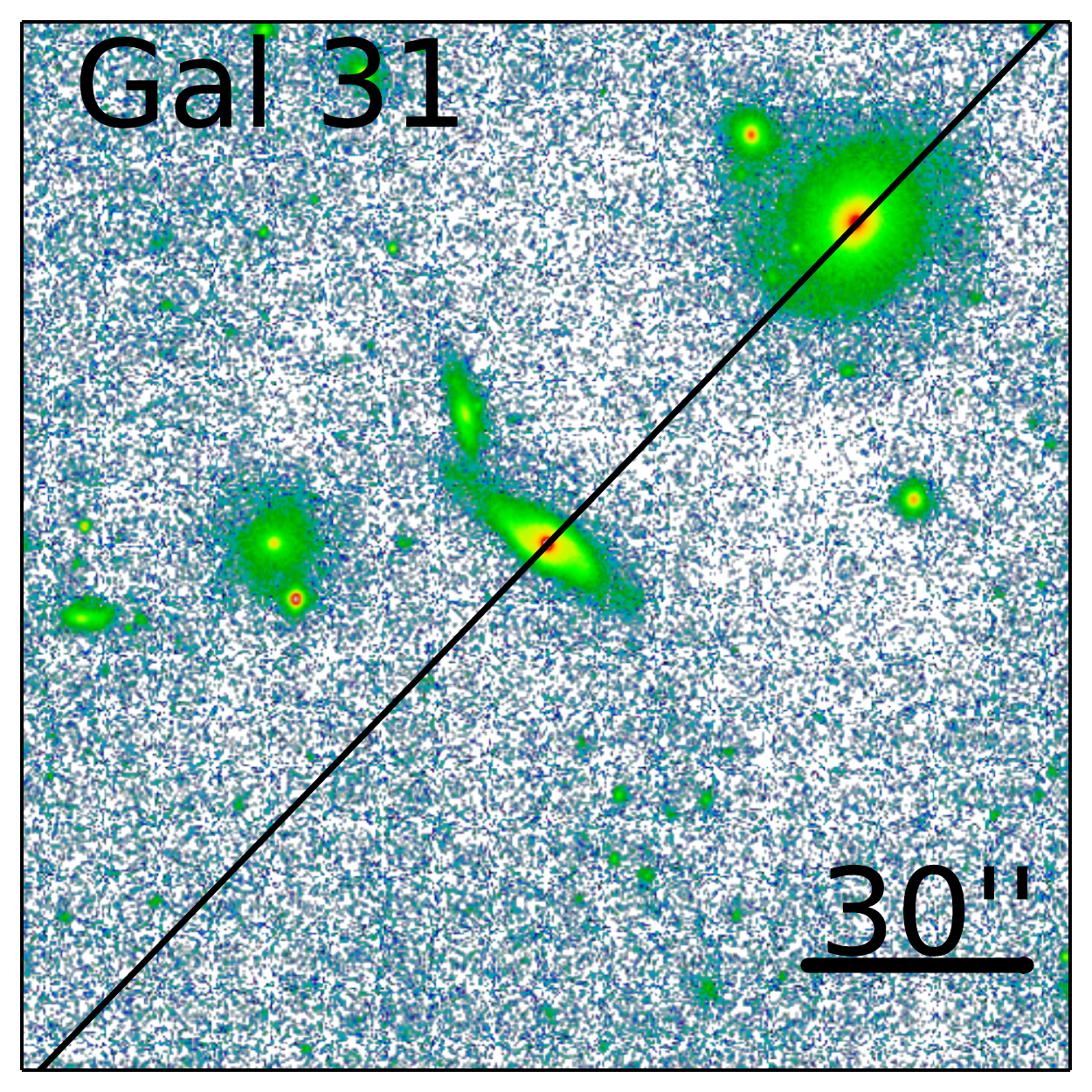}
\includegraphics[width=0.3\columnwidth]{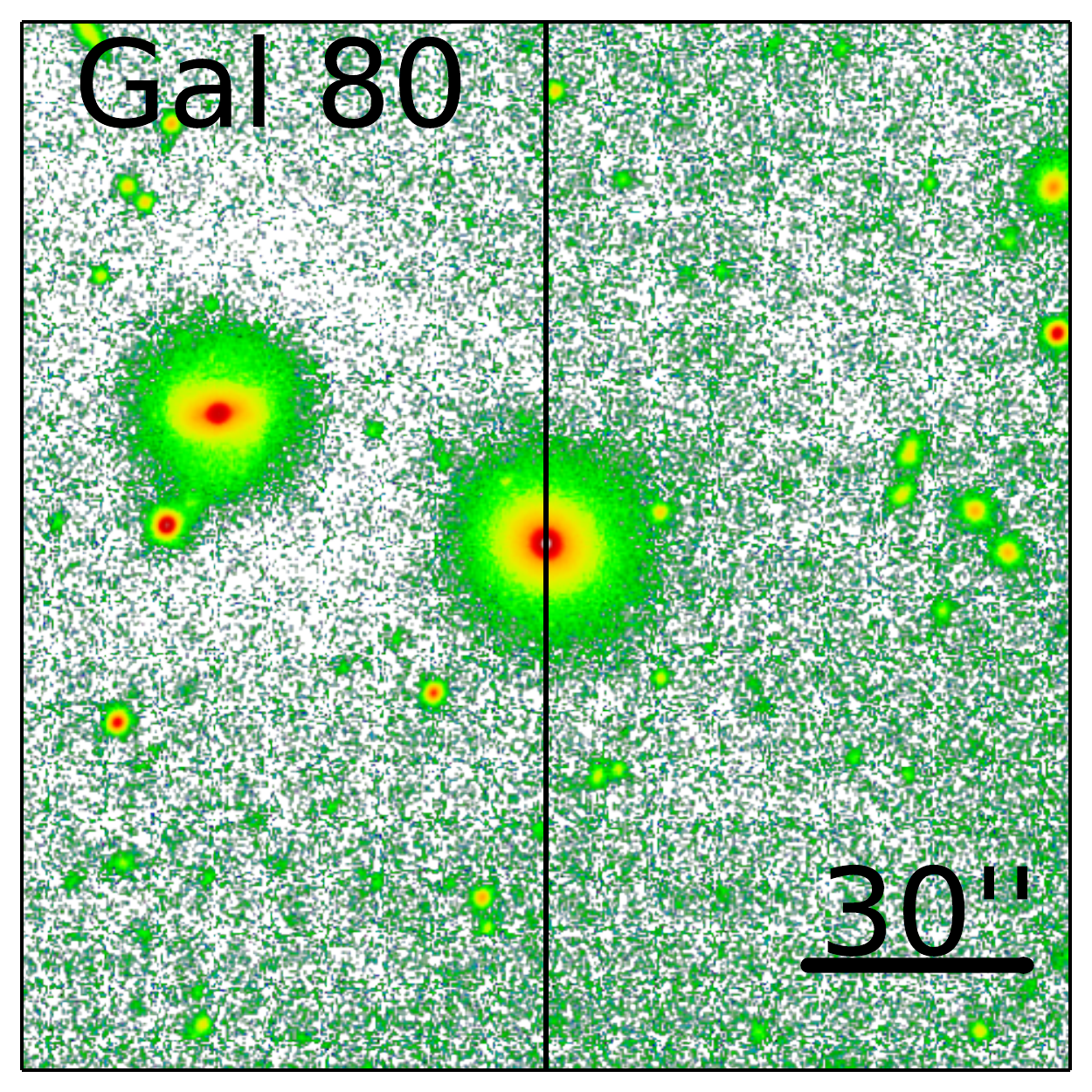}

\includegraphics[width=0.3\columnwidth]{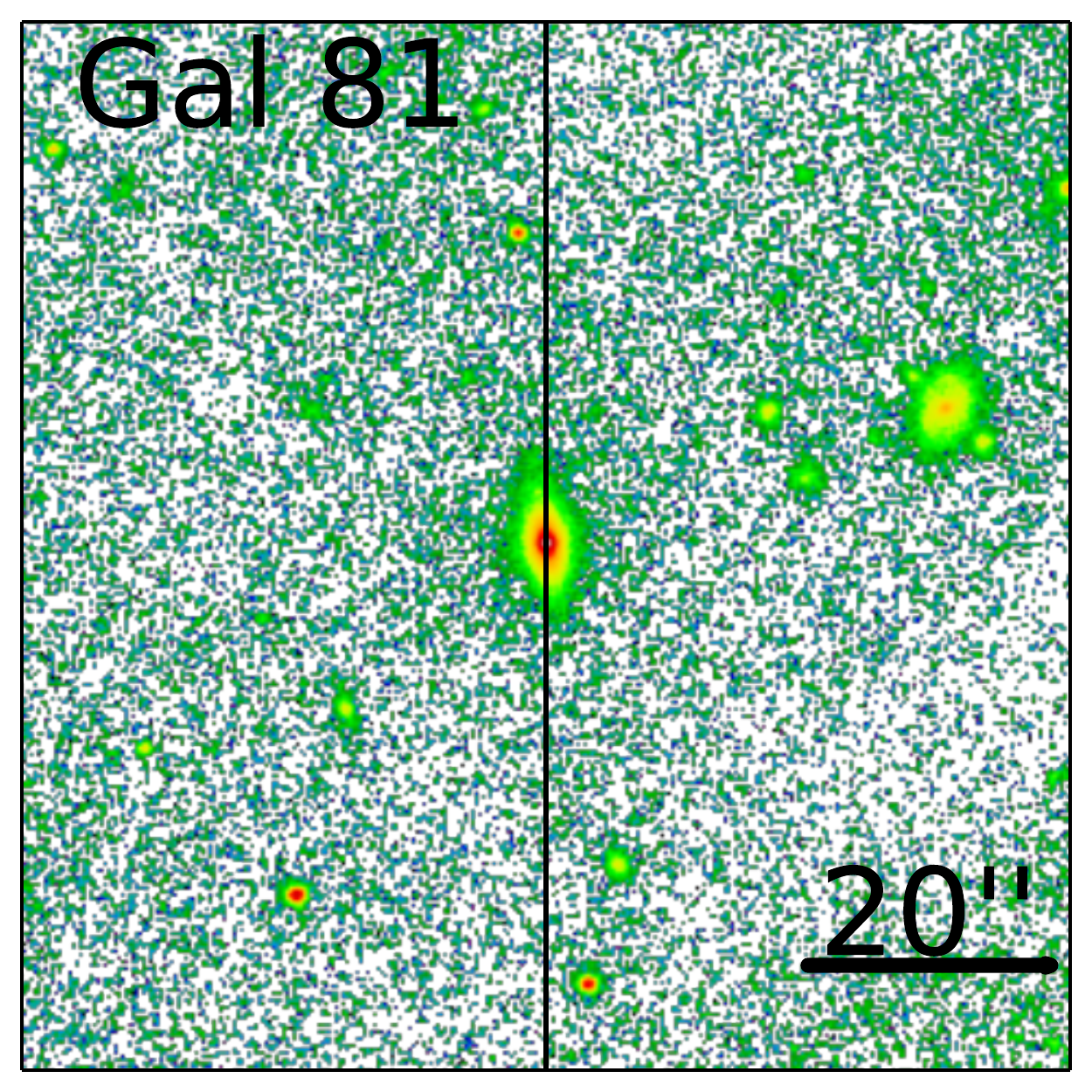}
\includegraphics[width=0.3\columnwidth]{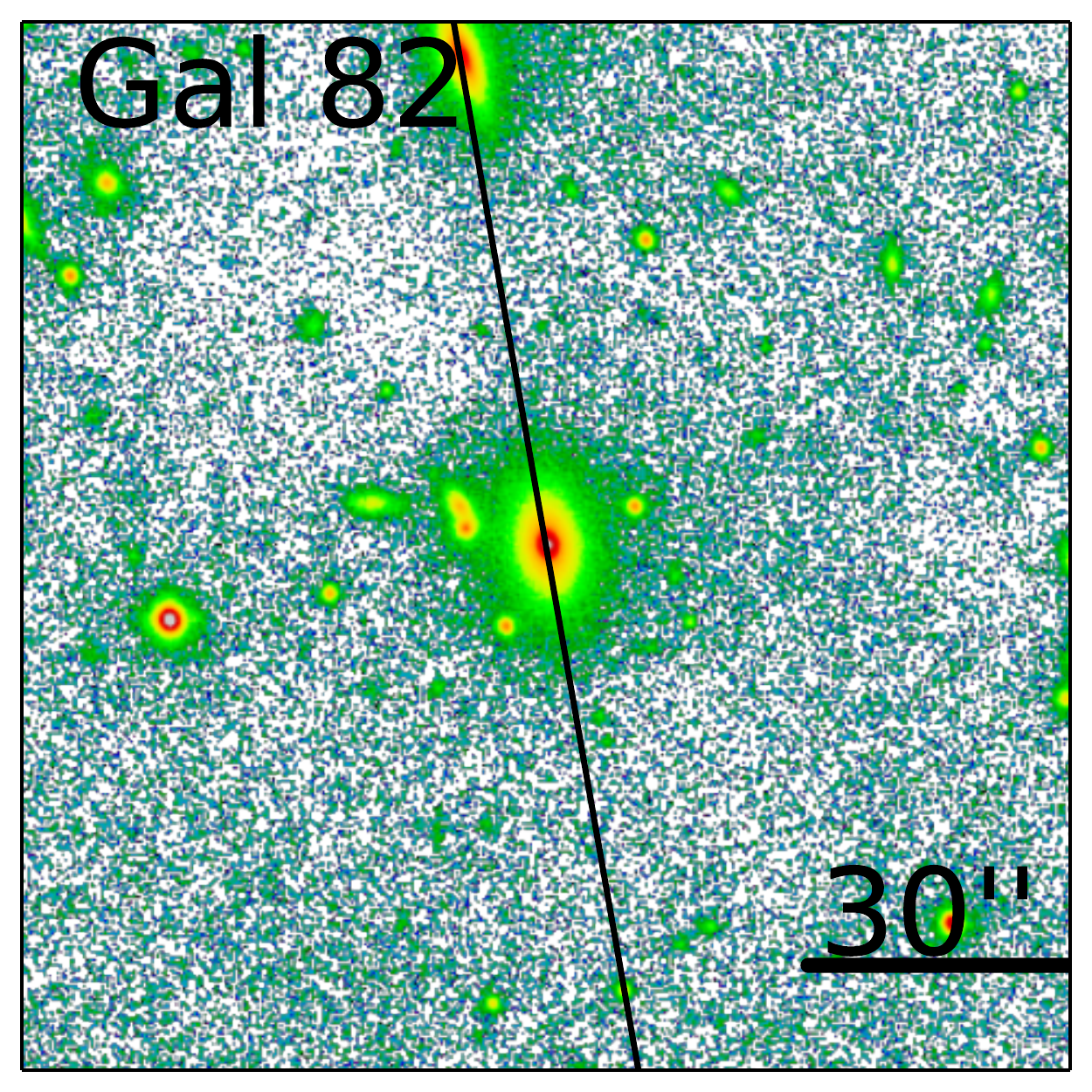}
\includegraphics[width=0.3\columnwidth]{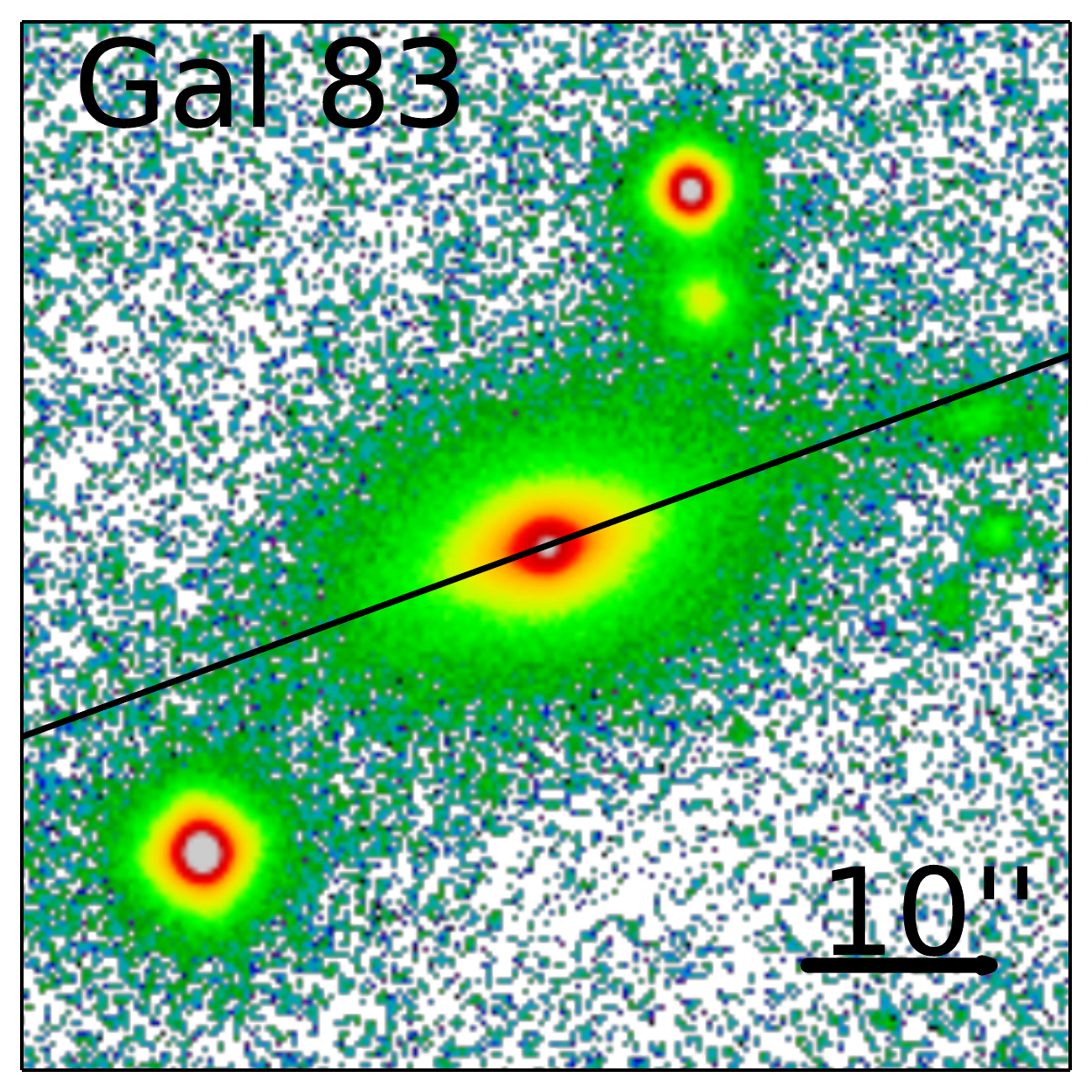}

\includegraphics[width=0.3\columnwidth]{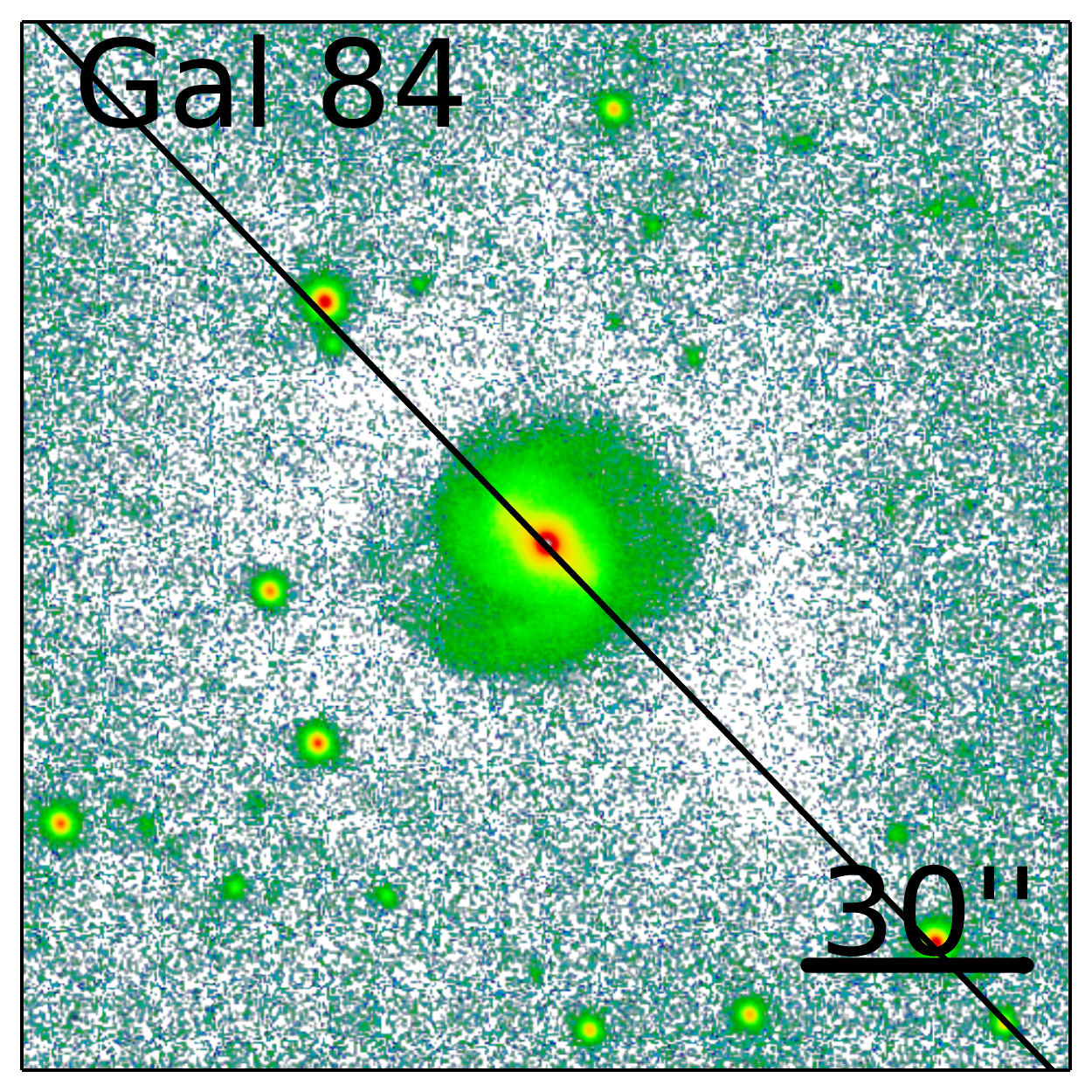}
\includegraphics[width=0.3\columnwidth]{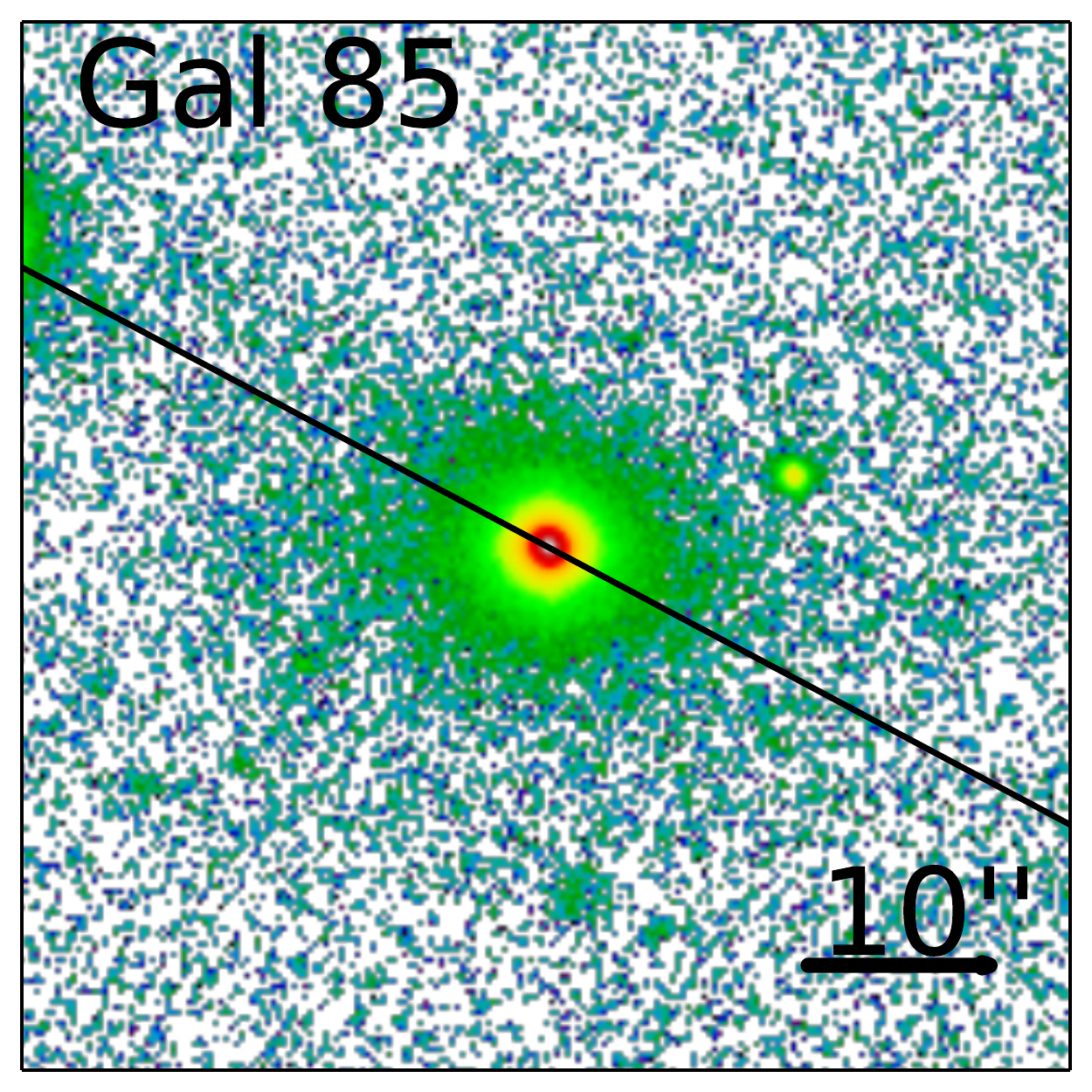}
\caption{SofI $H$-band images of the eleven analysed galaxies with slit position indicated. The slit width is 1\arcsec. North is up, East is left.}
\label{fig:slitplot}
\end{figure}

\begin{table*}
 \centering
  \caption{Source information: Coordinates, morphological classification, redshift, and observation parameters.}
  \begin{tabular}{ccccccccc}
  \hline \hline
source & RA (J2000) & DEC (J2000) & classification & redshift & $t_\mathrm{int}$ [s] & seeing [\arcsec] & PA slit [$^\circ$] & tell./calibr. star  \\
\hline
05 HE0036--5133 & 00h39m15.8s & -51d17m01s & E & 0.0288 & 7200 & 0.9 & 56 & HD 1839 \\
08 HE0045--2145 & 00h47m41.2s & -21d29m28s & SB & 0.0214 & 7200 & 1.5 & 25 & HD 224693 \\
11 HE0103--5842 & 01h05m16.7s & -58d26m14s & Irr & 0.0257 & 7200 & 1.4 & 112 & HD 14474 \\
16 HE0119--0118 & 01h21m59.8s & -01d02m24s & SB & 0.0547 & 7200 & 1.2 & 65 & HD 15760 \\
31 HE0323--4204 & 03h25m02.2s & -41d54m18s & S & 0.0580 & 8400 & 1.7 & 136 & HD 1839 \\
80 HE2112--5926 & 21h15m51.5s & -59d13m54s & E & 0.0317 & 7200 & 1.6 & 0 & HD 209552 \\
81 HE2128--0221 & 21h30m49.9s & -02d08m15s & E & 0.0528 & 7200 & 1.3 & 0 & HD 198273 \\
82 HE2129--3356 & 21h32m02.2s & -33d42m54s & E & 0.0293 & 7200 & 1.5 & 10 & HD 197612 \\
83 HE2204--3249 & 22h07m44.7s & -32d34m56s & E & 0.0594 & 6300 & 1.0 & 110 & HD 197612 \\
84 HE2211--3903 & 22h14m42.0s & -38d48m23s & SB & 0.0398 & 7200 & 1.0 & 44 & HD 221250 \\
85 HE2221--0221 & 22h23m49.5s & -02d06m13s & E & 0.0570 & 7200 & 1.3 & 62 & HD 198273 \\
\hline
\end{tabular}
\label{tab:sources}
\tablefoot{
The source designation consists of the ID from our LLQSO sample and the name from the Hamburg/ESO survey \citep[HE survey; ][]{2000A&A...358...77W}. Morphological classification is taken from \citet{2012nsgq.confE..60B}. The redshift is taken from the HE survey. The position angles are measured counter-clockwise from North to East.
}
\end{table*}

\section{Observation, reduction, and calibration}
\label{sec:obs}

Eleven galaxies from the \emph{low-luminosity type-1 QSO sample} have been observed in seeing limited mode with the Son of ISAAC (SofI) infrared spectrograph and imaging camera on the New Technology Telescope (NTT, ESO, Chile) during September 2009. The $1024\times 1024$ Hawaii HgCdTe array provides a pixel scale of $0.288\arcsec /\mathrm{pixel}$ with a field-of-view of $4.9\times4.9$ arcmin$^2$. \nnew{NIR imaging data are presented in \cite{2014A&A...561A.140B}.} The data analysed in this work consist of $H+K$-band low resolution longslit spectroscopy with a slit width of $1\arcsec$, resulting in a spectral resolution of about $R=\lambda/\Delta \lambda= 600$.

\nnew{For host galaxies showing prominent structures like bars, the slit was preferably aligned along those structures. In some cases, the slit position was chosen to include neighbouring objects. The positions of the slit superimposed on the $H$-band images from \cite{2014A&A...561A.140B} are shown in Fig.~\ref{fig:slitplot}.} Integration times, seeing conditions, and the position angles of the slit are presented in Table \ref{tab:sources}. The integration time was typically two hours and the seeing ranged from $0\farcs9$ to $1\farcs7$.

The spectra were taken with a nodding technique. Sky-subtraction was done by subtracting consecutive frames from each other. Flat-fielding, and the correction for tilt and curvature were done by \textsc{Iraf}/\textsc{Pyraf} standard procedures. Subsequently, the frames were shifted on top of each other and a median frame was calculated.

Telluric correction was performed with G2V-stars \nnew{which were observed before and after the science objects at similar airmass} (Table \ref{tab:sources}). The G2V class was chosen since they have the same spectral type as the Sun whose spectral lines are well known. For the telluric correction, the science spectra were divided by the telluric spectra (taking into account the different exposure times) and then multiplied by a solar reference spectrum \citep{1996AJ....111..537M} in order to correct for the black body shape and lines inserted by the telluric spectrum. 
\new{We shifted the spectra manually on subpixel-scale to match the absorption features and optimise the telluric correction. Since the solar reference spectrum was only available in the $H$- and $K$-band, we interpolated the region in between with a black body function with temperature $T=5800\,\mathrm{K}$. This means in the bandgap, we cannot correct for absorption lines of the G2V-type spectrum. Therefore, our telluric correction leaves residuals, especially in the bandgap between $H$- and $K$-band as well as in the region between $1.8 - 2\mm$ which has particularly low transmission. Line detections and measurements in this region have therefore to be taken with caution (however, the Pa$\alpha$-line that is located in the bandgap is quite strong and therefore well detected in most galaxies).}

\new{Wavelength calibration was done using the lines of a Xenon lamp observed every night. 
Flux calibration was done by scaling the solar reference spectrum to the 2MASS flux \citep{2006AJ....131.1163S} of the telluric star. In three nights, two or three standard stars were observed during the same night. In order to estimate the reliability of the flux calibration, we calibrated the standard stars with each other. The resulting calibrations varied by a factor of $10\% - 35\%$ likely caused by seeing variations during the night \citep[for a discussion of slit-loss corrections see][]{2015arXiv150901403R}.}

\section{Analysis and results}

In Figure \ref{fig:allspec}, we compare the central spectra which are extracted from an aperture with a diameter corresponding to the seeing FWHM (see Table \ref{tab:sources}) to off-nuclear spectra which are \nnew{extracted using the same aperture size for regions centred at $1.5\times$FWHM left and right from the nucleus.}

As expected, the continuum shapes of the off-nuclear spectra are bluer than those of the nuclear spectra which are much more affected by the hot dust emission from the obscuring torus. However, the difference in continuum shape between the eleven galaxies is much higher than the difference in continuum shape between different apertures. This hints at different excitation mechanisms and/or different impact of the central AGN emission on the galaxies' spectra that will be analysed in the following.

\begin{figure*}
\centering
\includegraphics[width=\linewidth]{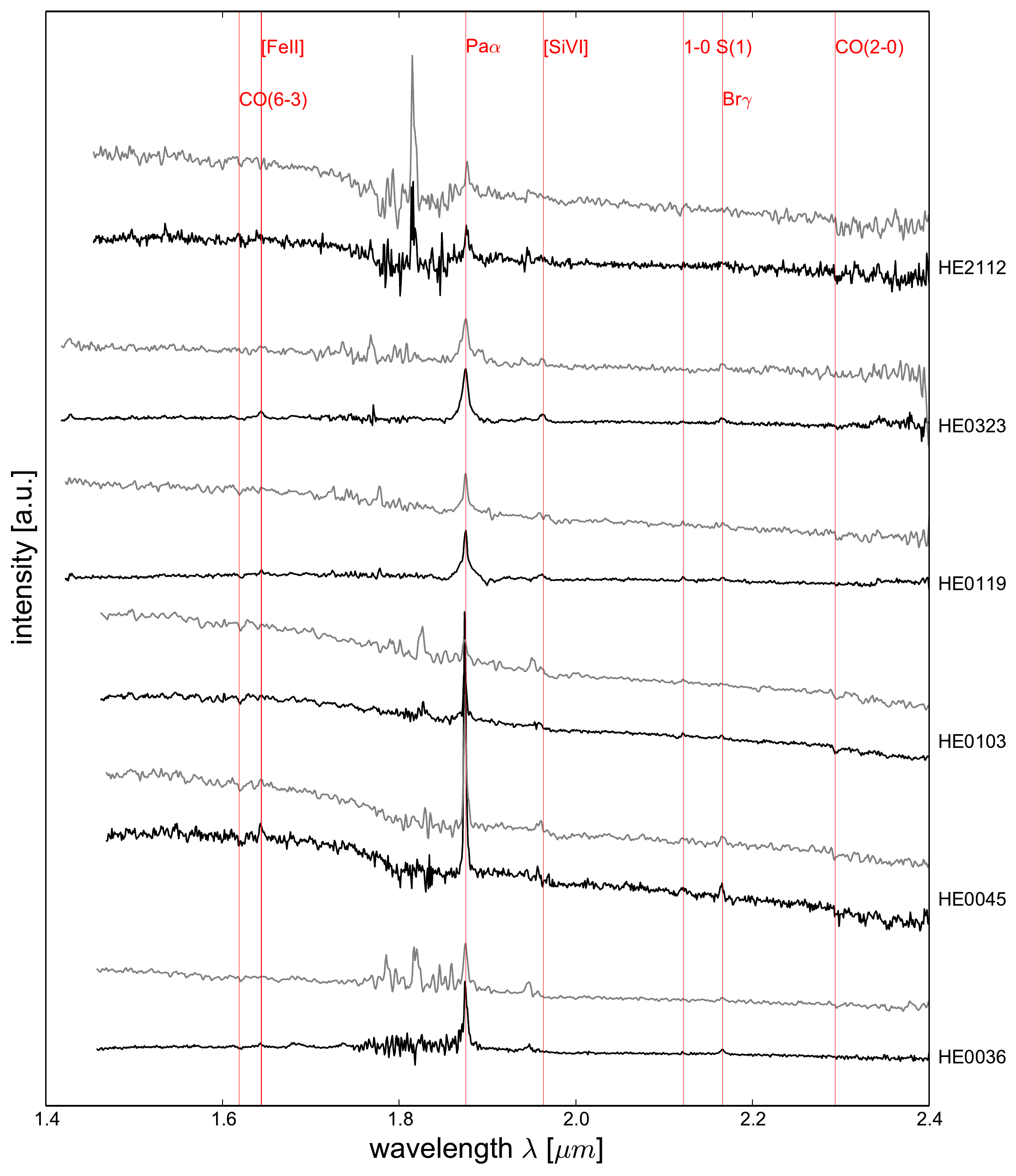}
\caption{$H+K$-band spectra of the objects analysed in this paper. For each galaxy, we show a central spectrum (black line/lower spectrum) which is extracted from an aperture with diameter corresponding to the FWHM of the seeing (see Table \ref{tab:sources}). In addition, we show an off-central spectrum (grey line/upper spectrum), centered 1.5$\times$FWHM left and right from the nucleus. The spectra are shown in restframe wavelength and have been normalized at 2.159$\mm$. Important emission and absorption lines have been marked.}
\label{fig:allspec}
\end{figure*}

\begin{figure*}
\ContinuedFloat
\centering
\includegraphics[width=\linewidth]{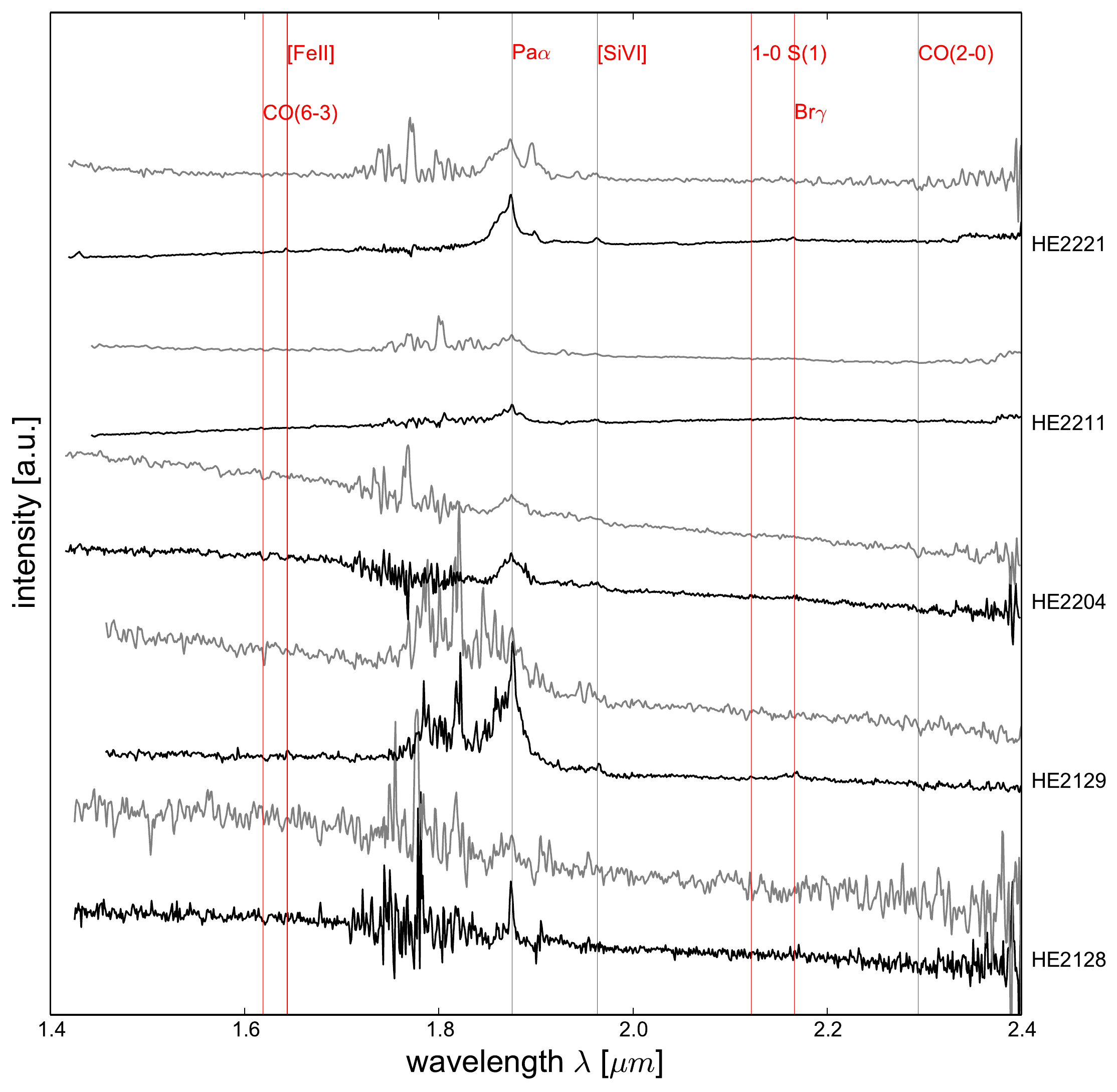}
\caption{continued}
\end{figure*}

\subsection{Continuum subtraction}
\label{sec:continuum}
\new{
The continuum emission of active galaxies is usually fitted by a multicomponent model, consisting of (1) an underlying power law, (2) thermal emission from hot dust, and (3) star light from the host galaxy. As pointed out in \cite{1987ApJ...320..537B}, the prominent bump in the NIR (peaking in the $K$-band at $\sim 2\mm$) can be produced by hot dust close to its sublimation temperature ($T\sim 1500\,\mathrm{K}$) that is predicted from the Unified Model \citep{1993ARA&A..31..473A,1995PASP..107..803U}. The power law emission is believed to originate from the accretion disk \citep[e.g.][]{1982ApJ...254...22M,1983ApJ...268..582M}. While it is prominent in the optical/UV, it becomes negligible in the NIR for Seyfert-2s \citep{2009ApJ...694.1379R}. However, in type-1 sources there might still be a significant contribution \citep[e.g.][]{2009MNRAS.400..273R,2011MNRAS.414..218L}. \cite{2006ApJ...640..579G} construct a spectral template for quasars from observations of 27 sources. In this template, the continuum emission in the NIR is best fitted by a combination of a power law ($f_\nu \propto \nu^\alpha$ with power-law index $\alpha=-0.92$) and a black body with temperature $T=1260\,\mathrm{K}$.
}

In the following, we fit the continuum shape, using the method described in \cite{2012A&A...544A.105S}. The continuum subtracted spectra are then used for emission line fits. Furthermore, the different fractions of the fitted functions give indication which processes contribute to the emission of the galaxies.

In the continuum fit, we include three components: hot dust, star light and a power law. Additionally, we consider extinction \citep{1989ApJ...345..245C}
\begin{equation}
\tau (\lambda,A_V) = - \frac{A_V}{1.086} \, \left[ \frac{0.574}{\lambda^{1.61}} - \frac{0.527}{\lambda^{1.61} R_V} \right]
\label{eq:extinction}
\end{equation}
with the standard value $R_V=3.1$ and $\lambda$ in $\mu m$. The hot dust contribution is modelled as black body radiation with the Planck function
\begin{equation}
\textrm{BB}(s_\mathrm{dust},\lambda,T) = s_\textrm{dust} \, \frac{2hc^2}{\lambda^5} \, \frac{1}{\exp \left( \frac{hc}{\lambda k T} \right) -1}
\label{eq:blackbody}
\end{equation}
with a scale factor $s_\textrm{dust}$. The contribution of the accretion disk is modelled by a power law
\begin{equation}
\textrm{PL}(s_\mathrm{AGN},\lambda,\alpha) = s_\textrm{AGN} \, \lambda^{-\alpha - 2}
\label{eq:powerlaw}
\end{equation}
with the power-law index $\alpha$ and the scale factor $s_\textrm{AGN}$. 
\new{
We fixed the power law index to the value from the quasar template of \cite{2006ApJ...640..579G} ($\alpha=-0.92$).
}
To account for the stellar contribution, we used stellar templates from the NASA Infrared Telescope Facility (IRTF) spectral library for cool stars \citep{2009ApJS..185..289R}. The library offers spectra of 210 cool stars (mainly F, G, K, M spectral type) at an resolving power of $\mathrm{R}\approx 2000$. The spectra are not normalized, i.e. the shape of the continuum is preserved.

The spectra had to be convolved with a Gaussian function in order to account for the lower resolution of our spectra and intrinsic effects such as broadening due to velocity dispersion or Doppler displacement:
\begin{equation}
\textrm{Star}(s_\textrm{star},\sigma, \Delta \lambda)= \textrm{Spec} * \frac{1}{\sigma \sqrt{2\pi}}  \exp \left( - \frac{(\lambda- \Delta \lambda)^2}{2\sigma^2} \right)
\label{eq:stellar}
\end{equation}

In total, we use the following function and fit it to the spectra using a \textsc{python} implementation of the \textsc{amoeba}-routine \citep[downhill simplex method,][]{1992nrfa.book.....P}:
\begin{multline}
\textrm{Cont}(A_V,s_\textrm{dust},T,s_\textrm{AGN},\alpha,s_\textrm{star},\sigma,\Delta \lambda) =
\exp \left( \tau (\lambda,A_V) \right) \times \\ \left[ \textrm{BB}(s_\mathrm{dust},\lambda,T) + \textrm{PL}(s_\mathrm{AGN},\lambda,\alpha) + \textrm{Star}(s_\textrm{star},\sigma, \Delta \lambda) \right]
\label{eq:fitfunction}
\end{multline}

The function has \new{six} fitting parameters: visual extinction $A_V$, temperature of the dust $T$, width $\sigma$ and displacement $\Delta \lambda$ of the Gaussian function \nnew{used to convolve the stellar templates, and the scale factors $s_\textrm{dust}$, $s_\textrm{AGN}$ and $s_\textrm{star}$ of the dust, power law, and stellar continuum}. 

\begin{figure}
\centering
\includegraphics[width=\columnwidth]{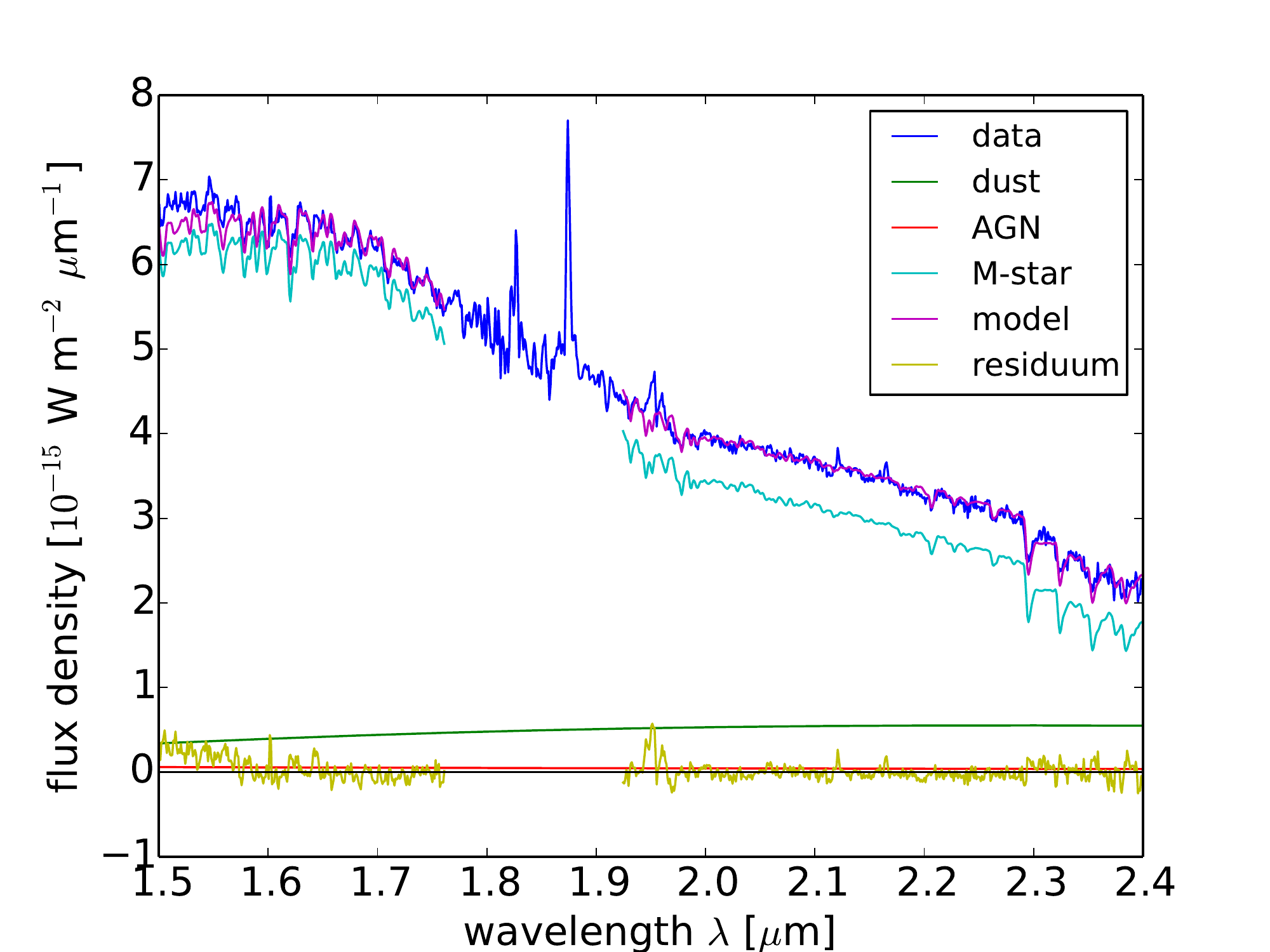}
\caption{Continuum fit. HE0103--5842 is shown as an example.}
\label{fig:contfit}
\end{figure}

One example of a continuum fit is displayed in Fig.~\ref{fig:contfit}. Table \ref{tab:continuum} shows the results of our continuum fits, i.e. the relative contributions of the contributing functions, as well as extinction, and dust temperature.
The fits give a first estimation of the contributing components and moreover, they are \nnew{used for continuum subtraction before fitting the emission lines}.
We point out that the fit might not always have a numerically unique solution
\new{
since dust emission ($T\approx 1200\,\mathrm{K}$) and strong extinction will both produce a red shape, while stellar emission and power law will both produce a blue spectrum. As pointed out by e.g. \cite{2004MNRAS.355..273C} and \cite{2009MNRAS.400..273R} young reddened starbursts are not distinguishable from a power law. Generally, for a more detailed analysis of the stellar content, a larger set of stellar populations is necessary. Using only one star as template often results in lower values for the stellar velocity dispersion \citep[see e.g.][]{2008MNRAS.385.1129R,2012A&A...544A.129V}. However, stellar kinematics or stellar population synthesis is not the goal of this work. The only aim is to subtract the continuum emission and get first estimates on the contributions of the different components. Our methods serves this purpose despite the mentioned limitations and degeneracies. 
}

\new{
In all galaxies (with exception HE0045-2145, see discussion below), we fitted a black body function with temperatures $1000\,\mathrm{K}\leq T_\mathrm{dust} \leq 1400\,\mathrm{K}$ that are typical for Seyfert-1 galaxies \citep[][and references therein]{2009MNRAS.400..273R,2010MNRAS.404..166R,2011MNRAS.414..218L}. In some galaxies, also a power-law component was fitted. All spectra show stellar components which contribute from 30\% up to almost 90\%. 
}
This shows the diversity of objects in terms of dominance of the nuclear component in the LLQSO sample.
\new{
We point out that the resulting stellar fractions are consistent with the AGN/host fractions derived from the decomposition of the NIR imaging data \citep{2014A&A...561A.140B} and with the amount of dilution in the CO band heads \citep[see e.g.][]{2007ApJ...671.1388D,2009MNRAS.400..273R} observed in the $K$-band spectra \nnew{(i.e. galaxies with dominant stellar component show prominent CO band heads, while the band heads are completely diluted by the underlying black body and/or power law in galaxies which are dominated by the AGN).}
}

\begin{table}
\centering
\caption{Results of the continuum fits: fraction of dust component (black body radiation) and its temperature $T$, fraction of the AGN component (power law), fraction of the stellar component, and the extinction.}
\label{tab:continuum}
\begin{tabular}{ccccccc} \hline \hline
source & dust & AGN & stars & $A_V$ & T(dust)  \\ \hline
05 HE0036 & 31\% & 33\% & 36\% & 0.8 mag & 1000 K \\
08 HE0045 & 32\% & 0\% & 68\% & 1.2 mag & 25000 K  \\
11 HE0103 & 13\% & 7\% & 80\% & 0 mag & 1400 K  \\
16 HE0119 & 55\% & 0\% & 45\% & 0 mag & 1300 K \\
31 HE0323 & 45\% & 20\% & 35\% & 0 mag & 1100 K  \\
80 HE2112 & 21\% & 0\% & 79\% & 0.6 mag & 1300 K \\
81 HE2128 & 18\% & 0\% & 82\% & 1.4 mag & 1200 K \\
82 HE2129 & 25\% & 0\% & 75\% & 0 mag & 1000 K \\
83 HE2204 & 10\% & 3\% & 87\% & 0.9 mag & 1000 K \\
84 HE2211 & 64\% & 4\% & 31\% & 0 mag & 1100 K \\
85 HE2221 & 59\% & 8\% & 33\% & 1.6 mag & 1000 K \\
\hline
\end{tabular}
\end{table}

\subsection{Emission line analysis}
\label{sec:emlines}

The strong Pa$\alpha$ hydrogen recombination line ($\lambda_\mathrm{rest}=1.87561\mm$) is located in the band gap between the $H$ and $K$-band which is \nnew{significantly affected by residuals from the telluric absorption}. Due to the redshift of the LLQSO sample ($0.02\leq z \leq 0.06$), the emission line is shifted towards the $K$-band in a region where it is still affected by the telluric absorption but already well detectable. In nine galaxies, the Pa$\alpha$ line could be separated into two components, a broad one (with FWHM typically between $2000$ and $5000\kms$) and a narrow component (FWHM below $1000 \kms$).

A second strong hydrogen recombination line located in the $K$-band is Br$\gamma$ ($\lambda_\mathrm{rest}=2.166112\mm$). Since the Br$\gamma$ line is usually weaker than Pa$\alpha$ by a factor of more than 10, we constrained the widths of Br$\gamma$ to the widths resulting from the Pa$\alpha$ fits. This reduces the number of fit parameters and makes the fits more robust.

All emission line fits have been performed using the \textsc{Python}-version of \textsc{Mpfitexpr} \citep{2009ASPC..411..251M}. 
\new{
We perform a Monte Carlo simulation with 100 iterations to estimate the uncertainties of the fit. In each iteration, we add Gaussian noise to the spectrum. The width of the noise distribution corresponds to the root-mean-square of the residual spectrum that we obtain after subtracting the fitted line from the spectrum. \nnew{This gives us the uncertainty at the position of the line, which is particularly useful in the case of the Pa$\alpha$ emission line that is located in a very noisy region of the spectrum.} In the following, we take the mean of the 100 fits as best fit and the standard deviation as uncertainty.
}
The results of the line fits are presented in Table \ref{tab:paabrgfits}.

\begin{table*}
 \centering
  \caption{Hydrogen recombination line fits. All fitted spectra have been extracted from apertures with diameter corresponding to $3\times$FWHM.}
  \begin{tabular}{ccccccc}
  \hline \hline
source & \multicolumn{2}{c}{Pa$\alpha$ broad} & \multicolumn{2}{c}{Pa$\alpha$ narrow} & {Br$\gamma$ broad} & {Br$\gamma$ narrow}  \\
 & flux & FWHM & flux & FWHM & flux & flux \\
 & [$10^{-20}\,\mathrm{W}\,\mathrm{m}^{-2}$] & [km s$^{-1}$] & [$10^{-20}\,\mathrm{W}\,\mathrm{m}^{-2}$] & [km s$^{-1}$] & [$10^{-20}\,\mathrm{W}\,\mathrm{m}^{-2}$] & [$10^{-20}\,\mathrm{W}\,\mathrm{m}^{-2}$] \\
\hline
05 HE0036--5133 & $1500\pm 200$ & 1300 & $480\pm 200$ & 260 & $160\pm 15$ & $27\pm5$ \\
08 HE0045--2145 & --- & --- & $1860\pm 50$ & 520 & --- & $130\pm 10$  \\
11 HE0103--5842 & --- & --- & $1140\pm 70$ & 610 & --- & $65\pm 10$ \\
16 HE0119--0118 & $3500\pm 120$ & 3200 & $1080\pm 60$ & 560 & $280\pm 140$ & $110\pm 20$  \\
31 HE0323--4204 & $2000\pm 110$ & 3200 & $1080\pm 150$ & 900 & $120\pm 70$ & $170\pm 30$ \\
80 HE2112--5926 & $1300\pm 300$ & 2900 & $380\pm 230$ & 470 & --- & --- \\
81 HE2128--0221 & $180\pm 90$ & 1800 & $200\pm 70$ & 620 & --- & --- \\
82 HE2129--3356 & $7500\pm 600$ & 5000 & $1150\pm 420$ & 720 & $500\pm 80$ & $70\pm 20$ \\
83 HE2204--3249 & $2900\pm 120$ & 4000 & $80\pm 30$ & 250 & $410\pm 60$ & $24\pm 13$ \\
84 HE2211--3903 & $6400\pm 300$ & 3900 & $390\pm 80$ & 420 & $1400\pm 80$ & $50\pm20$ \\
85 HE2221--0221 & $18800\pm 300$ & 4300 & $2100\pm 100$ & 730 & $1600\pm 130$ & $150\pm 40$ \\
\hline
\end{tabular}
\label{tab:paabrgfits}
\end{table*}


More than half of the galaxies in the sample show significant emission in molecular hydrogen. Particularly the transitions H$_2$(1-0)S(3) ($\lambda 1.958\mm$), H$_2$(1-0)S(2) ($\lambda 2.034\mm$), and H$_2$(1-0)S(1) ($\lambda 2.122\mm$) are detected. \nnew{This shows that LLQSOs are not only rich in cold molecular gas \citep[see][]{2007A&A...470..571B}, but also contain warm, excited molecular gas.} Ratios of molecular hydrogen lines can give valuable information on the excitation mechanisms \citep{2003ApJ...597..907D,2005ApJ...633..105D}. The observations analysed here are too shallow to investigate H$_2$ excitation mechanisms in detail. Nevertheless, the detection of molecular hydrogen in this study helps to select candidates for follow-up studies on the hydrogen excitation mechanisms based on deeper observations. The results of the emission line fits are reported in Table \ref{tab:h2feiifits}.

Most of the galaxies show emission in the forbidden iron transition [\ion{Fe}{ii}] at $\lambda 1.644\mm$. This line can be excited by the AGN but is also known as shock tracer and particularly used as supernova rate estimator \citep[e.g.][]{1997AJ....113..162C,2003AJ....125.1210A}. Furthermore, some galaxies, particularly HE2129--3356 and HE2221-0221, show strong emission in the coronal line [\ion{Si}{vi}] which is a common AGN tracer.

\new{
In their studies of Seyfert galaxies and quasars, \citet{riffel_0.8-2.4_2006} and \citet{mason_nuclear_2015} detect the same set of lines. They report that [\ion{Fe}{ii}], H$_2$, and [\ion{Si}{vi}] are detected more frequently in nearby AGN than in quasars. Our detection rates in low-luminosity QSOs ([\ion{Fe}{ii}]: 7/11, H$_2$(1-0)S(1): 6/11, see Table \ref{tab:h2feiifits}) lie between their detection rates for AGN and quasars, as expected for a ``bridge'' sample between these two populations.
}

\begin{table*}
 \centering
  \caption{Emission line fluxes of molecular hydrogen lines and of the forbidden line [\ion{Fe}{ii}]. All fitted spectra have been extracted from apertures with diameter corresponding to $3\times$FWHM.}
  \begin{tabular}{ccccccc}
  \hline \hline
source & H$_2$(1-0)S(3) & H$_2$(1-0)S(2) & H$_2$(1-0)S(1) & [\ion{Fe}{ii}] \\
 & $\lambda 1.958\mm$ & $\lambda 2.034\mm$ & $\lambda 2.122\mm$ & $\lambda 1.644\mm$ \\
 & [$10^{-20}\mathrm{W}\,\mathrm{m}^{-2}$] & [$10^{-20}\mathrm{W}\,\mathrm{m}^{-2}$] & [$10^{-20}\mathrm{W}\,\mathrm{m}^{-2}$] & [$10^{-20}\mathrm{W}\,\mathrm{m}^{-2}$] \\
\hline
05 HE0036--5133 & $58\pm 12$ & --- & $31\pm8$ & $200\pm 30$ \\
08 HE0045--2145 & $110\pm 20$ & --- & $120\pm40$ & $180\pm 20$ \\
11 HE0103--5842 & $130\pm 40$ & --- & $90\pm 30$ & $170\pm 20$ \\
16 HE0119--0118 & $170\pm 40$ & $50\pm 20$ & $140\pm 20$ & $190\pm 20$ \\
31 HE0323--4204 & $300\pm 20$ & --- & $50\pm 30$ & $300\pm 40$ \\
80 HE2112--5926 & --- & --- & --- & --- \\
81 HE2128--0221 & --- & --- & --- & --- \\
82 HE2129--3356 & $200\pm 80$ & --- & $60\pm 20$ & $130\pm 30$ \\
83 HE2204--3249 & --- & --- & --- & --- \\
84 HE2211--3903 & --- & --- & --- & --- \\
85 HE2221--0221 & --- & --- & --- & $370\pm 30$ \\
\hline
\end{tabular}
\label{tab:h2feiifits}
\end{table*}

\section{Discussion}
\subsection{Excitation mechanisms}
\label{sec:excitation}

In the optical, emission line ratios are commonly used to distinguish between different excitation mechanisms (mainly star formation vs. AGN ionization) making use of different \emph{diagnostic diagrams} \citep[e.g.][]{1981PASP...93....5B,2001ApJ...556..121K,2003MNRAS.346.1055K,2007MNRAS.382.1415S,2013A&A...558A..34B,2015A&A...573A..93V}.

A diagnostic diagram in the NIR has been \new{suggested by \citet{1998ApJS..114...59L} and further developed by \citet{2004A&A...425..457R,2005MNRAS.364.1041R,2013MNRAS.430.2002R}.} It uses the line ratios $\log(\mathrm{H}_2(1-0)\mathrm{S}(1)/\mathrm{Br}\gamma)$ and $\log ([$\ion{Fe}{ii}$]\lambda 1.257\mm/\mathrm{Pa}\beta)$ to distinguish between excitation from pure photoionization and from pure shocks. It shows a transition from starburst galaxies to LINERs, passing AGNs where both excitation mechanisms are of importance \citep[see also models by][]{2012MNRAS.422..252D}. We use the conversion factors $[$\ion{Fe}{ii}$]\lambda 1.644\mm/[$\ion{Fe}{ii}$]\lambda 1.257\mm=0.744$ \citep{1988A&A...193..327N} and $\mathrm{Pa}\alpha/\mathrm{Pa}\beta=2.05$ \citep{2006agna.book.....O} to convert the accessible Pa$\alpha$ and $[$\ion{Fe}{ii}$]\lambda 1.644\mm$ line fluxes to the needed $J$-band line fluxes. 
\new{
In contrast to the $J$-band ratio, the used lines Pa$\alpha$ and $[$\ion{Fe}{ii}$]\lambda 1.644\mm$ are seperated in wavelength by an amount where reddening effects become significant. The line ratio will change by a factor of $10^{0.4\times [ A([\ion{Fe}{ii}]) - A(\mathrm{Pa}\alpha) ]}$ which corresponds to $10^{0.0204\times A_V}$ for the \citet{2000ApJ...533..682C} extinction law ($A([\ion{Fe}{ii}])$ and $A(\mathrm{Pa}\alpha)$ denote the extinction at the wavelength of [\ion{Fe}{ii}] and Pa$\alpha$ resp., while $A_V$ denotes visual extinction). This means that for a typical extinction of $A_V=2$, we underestimate the line ratio by 10\% and by 60\% for an extreme extinction of $A_V=10$. Since extinction could not be reliably determined from our data, we include this uncertainty in the error bars.
}

Nine out of eleven observed galaxies have the required line detections to be placed in the diagnostic diagram at least with upper limits. All of these galaxies are consistent with AGN photoionization. 
\new{This confirms that LLQSOs already show a significant to dominating contribution of non-stellar emission from the central AGN. However, some galaxies are shifted towards the region where starburst dominated galaxies are located. This underlines the importance of both ionization mechanisms in the LLQSO sample. The more detailed analysis of individual objects (Appendix \ref{sec:individual}) shows more indications for ongoing star formation in several objects.

With the present data, we can only probe the excitation mechanisms in a single aperture, corresponding to the central $\sim 1\,\mathrm{kpc}$. With follow-up integral-field spectroscopy, we will be able to trace the ionization mechanisms in a spatially resolved way \citep[see pilot study by][]{2015A&A...575A.128B}. 
}

\begin{figure}
\centering
\includegraphics[width=\columnwidth]{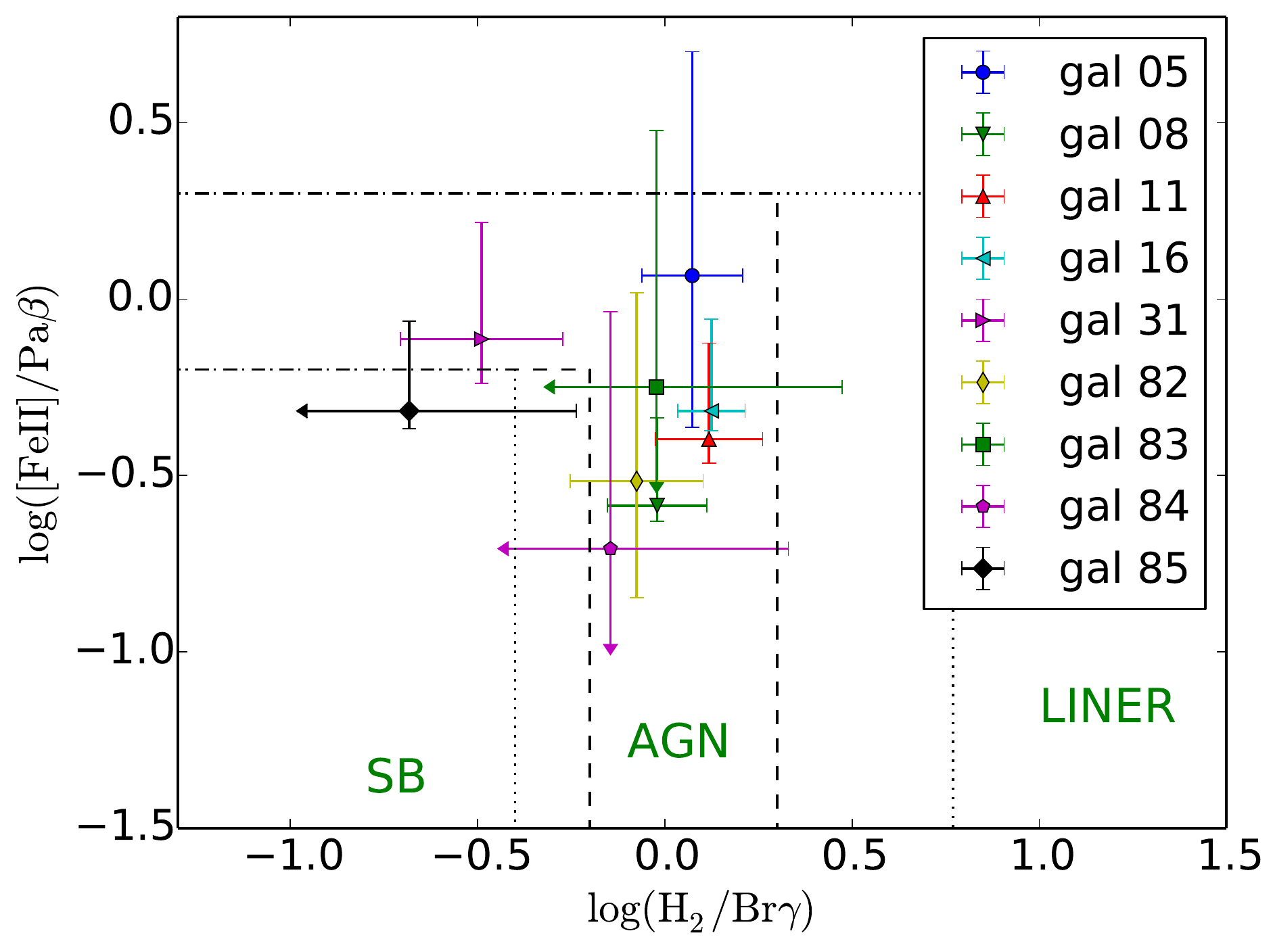}
\caption{Diagnostic diagram in the NIR with line ratios $\log(\mathrm{H}_2(1-0)\mathrm{S}(1)/\mathrm{Br}\gamma)$ and $\log ([$\ion{Fe}{ii}$]\lambda 1.257\mm/\mathrm{Pa}\beta)$ (for details see text). Apertures have a diameter corresponding to $3\times$FWHM of the seeing. The positions of 9 galaxies in the diagnostic diagram are shown. The lines indicate regions that are typically populated by starburst galaxies, AGNs, and LINERs resp. Dashed lines: \citet{2005MNRAS.364.1041R} and \citet{2010MNRAS.404..166R}, dotted lines: \citet{2013MNRAS.430.2002R}.}
\label{fig:diagndiagram}
\end{figure}

\subsection{Black hole mass - bulge luminosity relation}
\label{sec:bhmass}

We use the broad component of the Pa$\alpha$ emission line that we detect in nine out of eleven galaxies to estimate the mass of the central supermassive black hole. According to \cite{2010ApJ...724..386K}
\begin{equation}
M_\textrm{BH} = 10^{7.16\pm 0.04} \left( \frac{L_{\mathrm{Pa}\alpha}}{10^{35} \mathrm{W}} \right)^{0.49\pm 0.06} \left( \frac{\mathrm{FWHM}_{\mathrm{Pa}\alpha}}{10^3\  \mathrm{km}\ \mathrm{s}^{-1}} \right)^2 M_\odot.
\label{eq:massbh}
\end{equation}
 The calculated black hole masses are presented in Table \ref{tab:bhmasses}. We compare them to black hole masses available from \cite{2010A&A...516A..87S} and Schulze (priv. communication). They used optical spectra and computed BH masses using the scaling relation between broad line region (BLR) size and continuum luminosity by \cite{2009ApJ...697..160B} and the scale factor $f=3/4$ from \cite{1990ApJ...353..108N} which results in:
\begin{equation}
M_\mathrm{BH} = 5.025 \times \left( \frac{L_{5100}}{10^{37}~ \mathrm{W}} \right)^{0.52} \left( \frac{\mathrm{FWHM}_{\mathrm{H}\beta}}{\mathrm{km}~\mathrm{s}^{-1}} \right)^2 M_\odot.
\end{equation}

The mass estimates agree well with each other. The deviations are $\leq 0.3$ dex which is small given the uncertainties of BH mass estimates in general. The BH mass estimates from Pa$\alpha$ in the NIR thus confirm Schulze's masses from H$\beta$ that we used in \citet{2014A&A...561A.140B} to test the $M_\mathrm{BH}-L_\mathrm{bulge}$ relation. 

\begin{figure}
\centering
\includegraphics[width=\columnwidth]{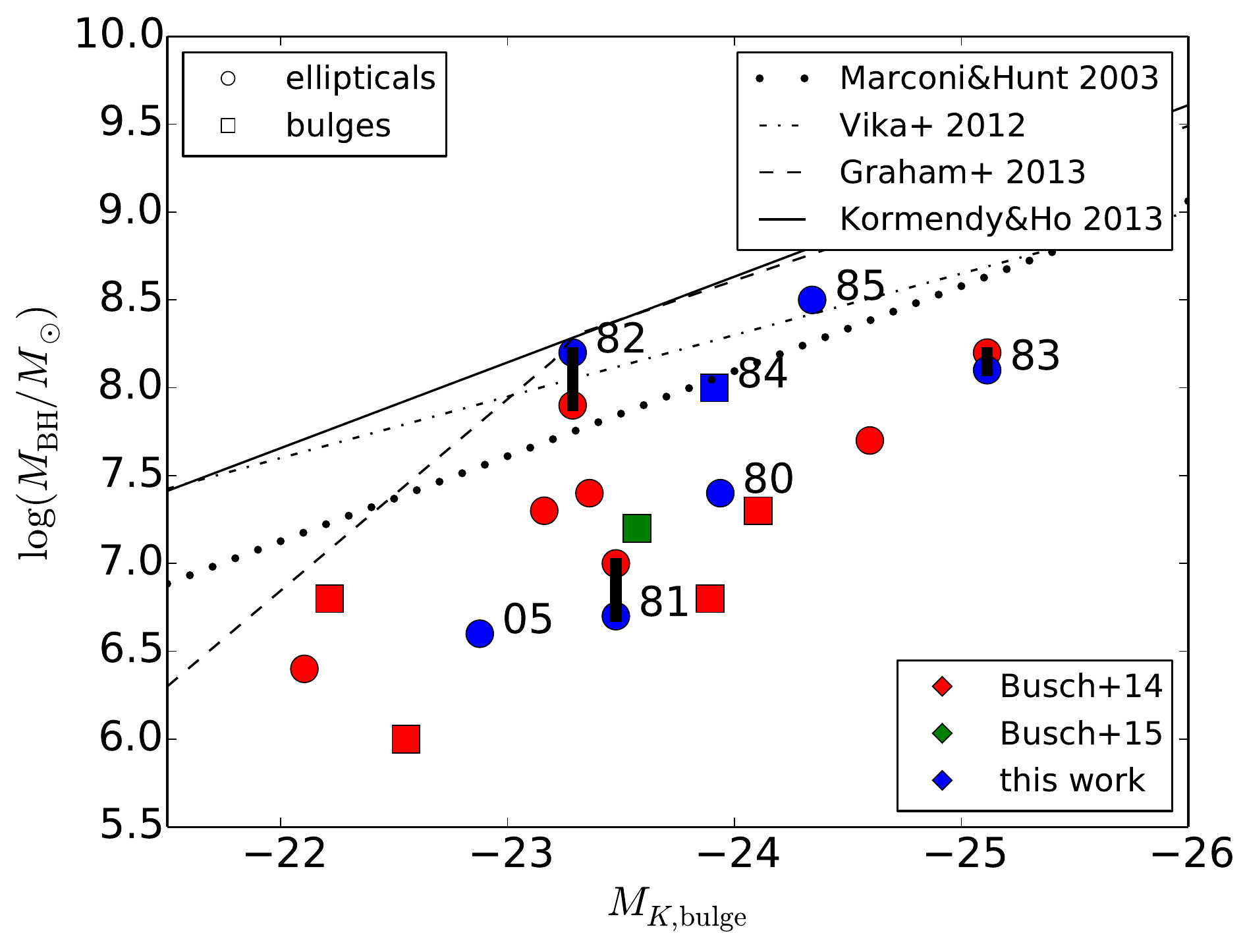}
\caption{Correlation between black hole mass $M_\mathrm{BH}$ and absolute $K$-band magnitude $M_K$ of the spheroidal component. Bulge magnitudes are from \citet{2014A&A...561A.140B}. Black hole masses for the red points are from A.~Schulze \citep[data points already published in][]{2014A&A...561A.140B}. The green point is taken from \citet{2015A&A...575A.128B}. Four data points with black hole masses derived from NIR spectroscopy in this work are added in blue.}
\label{fig:bhmassrel}
\end{figure}

Furthermore, we get BH masses for four galaxies that we did not have BH masses of before \nnew{(HE0036--5133, HE2112--5926, HE2211--3903, HE2221--0221)}. We use the $K$-band bulge magnitudes derived in \citet{2014A&A...561A.140B} to plot them in the $M_\mathrm{BH}-L_\mathrm{bulge}$ diagram (Fig.~\ref{fig:bhmassrel}). The new data points confirm the results found in the previous studies \citep{2014A&A...561A.140B,2015A&A...575A.128B}. With the new data we enlarged the sample from 12 to 16 LLQSOs that systematically lie below published $M_\mathrm{BH}-L_\mathrm{bulge}$ relations of inactive galaxies which makes the findings even more robust.

\new{
Recently, specific growth rates of the BH and the stellar host have been used to explore the time evolution (or ``flow patterns'') in the $M_\mathrm{BH}-M_*$ relation. AGN that are offset from the relation mostly have evolutionary vectors which are anti-correlated with their positions. That means, that they are moving back towards the relation \citep{2010ApJ...708..137M,2015ApJ...802...14S}.
}
As discussed in \citet{2014A&A...561A.140B}, possible explanations for a deviation of active galaxies from a near-linear $M_\mathrm{BH}-L_\mathrm{bulge}$ relation are (a) bulges which contain young stellar populations (as opposed to the usual picture that bulges are mainly consisting of old stars) and are therefore brighter at given mass, or (b) undermassive black holes that are in a growing phase. 
Both scenarios fit into a framework of bulge and black hole coevolution that is not solely based on a ``classical'' merging of galaxies but includes interaction of the black hole and the surrounding host galaxy via AGN feeding and feedback. The offset of the LLQSOs in the BH - bulge relation might thus shed light on the sequence of star forming and BH fuelling/feedback phases which is an important component for understanding the BH - bulge coevolution.

According to the NIR diagnostic diagram (Sect.~\ref{sec:excitation}) the analysed galaxies are dominated by the AGN. However, some objects are shifted towards the location of star formation dominated sources. Furthermore, in the discussion of individual objects (\new{see Appendix}) we point out that many objects show indications for ongoing star formation. \nnew{This could indicate overluminous bulges associated with lower mass-to-light ratios.}

\nnew{As pointed out in \citet{2007A&A...470..571B}, LLQSOs are rich in cold molecular gas. Here, we find more than half of our objects to show molecular hydrogen emission in the near-infrared, which indicates that part of the molecular gas content is also excited.} Furthermore, several of the objects can be classified as (ultra-)luminous infrared galaxies ((U)LIRGs). 
\nnew{
While the presence of molecular gas and star formation are related \citep[e.g. see the famous Kennicutt-Schmidt law;][]{1998ApJ...498..541K} and molecular gas is therefore a prerequisite for star formation, the role of molecular gas in the context of AGN fueling is not fully understood. On the one hand, \citet[][and references therein]{2015MNRAS.451.3587R} show that H$_2$ is usually located in a disk in the plane of the galaxy and often shows inflows to the center (is a tracer of AGN fueling). On the other hand, cold molecular gas can also be found in outflows \citep[feedback;][]{2014A&A...567A.125G}.
}

Recently, an alternative explanation for the offset in the $M_\mathrm{BH}-L_\mathrm{bulge}$ relation has been proposed \citep{2012ApJ...746..113G,2013ApJ...764..151G,2013ApJ...768...76S,2015ApJ...798...54G}. They suggest that the relation between BH mass and stellar mass (and luminosity) of the host spheroid is a ``broken'' relation, with a near-linear $M_\mathrm{BH}-M_{\mathrm{bulge},*}$ relation only for high BH masses ($M_\mathrm{BH}\gtrsim 10^8\,M_\odot$). Galaxies with lower BH masses are found to follow a different, steeper relation which indicates that the black hole is growing more rapidly than the surrounding spheroid. \citet{2015arXiv150102937G} states that the LLQSOs presented in \citet{2014A&A...561A.140B} follow the relation of lower-mass galaxies. We see that all eleven galaxies (plus the five galaxies added since then) are systematically shifted below the relation though. This indicates that LLQSOs are also offset from the new relation. \nnew{However, reliable mass-to-light ratios are indispensable to decide on the validity of the offset of active galaxies. As discussed above, a lower mass-to-light ratio would be expected in the case of enhanced nuclear star formation.}

By using integral-field spectroscopy, \citet{2015A&A...575A.128B} show that in the case of HE 1029--1831 circumnuclear star formation shifts the LLQSO away from the $M_\mathrm{BH}-L_\mathrm{bulge}$ relation of inactive galaxies. Only by using spatially resolved data at highest angular resolution (using-adaptive optics assisted integral-field spectroscopy with e.g. SINFONI at the Very Large Telescope) as done in the mentioned pilot study, central AGN emission and circumnuclear star formation can be reliably disentangled.

\begin{table*}
\centering
\caption{BH mass estimates from the broad Pa$\alpha$ line compared with A.~Schulze's BH mass estimates from $\mathrm H \alpha$. Bolometric luminosities are derived from ROSAT soft X-ray luminosities.}
\label{tab:bhmasses}
\begin{tabular}{c|ccc|ccc|c} \hline \hline
source & $\log(L_{\mathrm{Pa}\alpha}[\mathrm{W}]$ & $\mathrm{FWHM}_{\mathrm{Pa}\alpha}$ & $\log(M_\textrm{BH}/M_\odot)$ & $\log(L_{5100}[W])$ & $\mathrm{FWHM}_{\mathrm{H}\alpha}$ & $\log(M_\textrm{BH}/M_\odot)$ & $\log(L_\mathrm{bol}[\mathrm{W}])$ \\
 &  & [$\mathrm{km}\ \mathrm{s}^{-1}$] & & & [$\mathrm{km}\ \mathrm{s}^{-1}$] & \\
 \hline
05 HE0036--5133 & $33.5$ & 1300 & 6.6 & --- & --- & --- & 38.1\tablefootmark{a}  \\
08 HE0045--2145 & --- & --- & --- & $36.2$ & $682$ & 6.0 & --- \\
11 HE0103--5842 & --- & --- & --- & $36.2$ & $1729$ & 6.8 & 37.7 \\
16 HE0119--0118 & $34.4$ & 3200 & 7.9 & $36.8$ & $4192$ & 7.9 & 38.0  \\
31 HE0323--4204 & $34.2$ & 3200 & 7.8 & $36.8$ & $3030$ & 7.6 & 38.0 \\
80 HE2112--5926 & $33.5$ & 2900 & 7.4 & --- & --- & --- & --- \\
81 HE2128--0221 & $33.1$ & 1800 & 6.7 & $36.5$ & $1485$ & 7.0 & 37.7 \\
82 HE2129--3356 & $34.2$ & 5000 & 8.2 & $36.5$ & $5197$ & 7.9 & 38.5 \\
83 HE2204--3249 & $34.4$ & 4000 & 8.1 & $36.9$ & $5770$ & 8.2 & 38.7 \\
84 HE2211--3903 & $34.4$ & 3900 & 8.0 & --- & --- & --- & 38.0 \\
85 HE2221--0221 & $35.2$ & 4300 & 8.5 & --- & --- & --- & --- \\
\hline
\end{tabular}
\tablefoot{
The bolometric luminosities are derived from the unabsorbed ROSAT fluxes \citep{2010MNRAS.401.1151M}, using a bolometric correction factor of $\sim 50$ \citep{2007ApJ...654..731H}.
\tablefoottext{a}{HE0036 is an X-ray transient source \citep{1995A&A...300L..21G} and was at its peak luminosity during the ROSAT observations.}
}
\end{table*}

\subsection{Far-infrared properties}
\label{sec:firprop}

Seven galaxies are listed in the \emph{IRAS Faint Source Catalog} \citep{1990IRASF.C......0M}. We present the fluxes measured at $12\mm$, $25\mm$, $60\mm$, and $100\mm$ in Table \ref{tab:iras}. From these, infrared luminosities 
\begin{multline}
L_\mathrm{IR}(8-1000\mm)[\mathrm{W}] = 4\pi \left( D_L[\mathrm{Mpc}] \right)^2 \times 1.72 \times 10^{31} \\ \times \left(13.48\ F_{12\mm} + 5.16\ F_{25\mm} + 2.58\ F_{60\mm} + F_{100\mm} \right)
\end{multline}
and far-infrared luminosities
\begin{multline}
L_\mathrm{FIR}(40-500\mm)[\mathrm{W}] \\= 4\pi \left( D_L[\mathrm{Mpc}] \right)^2 \times 1.2 \times 10^{31} \left( 2.58\ F_{60\mm} + F_{100\mm} \right)
\end{multline}
can be calculated \citep{1996ARA&A..34..749S}. With the calibration of \cite{2003A&A...409...99P}, we can further calculate FIR-star formation rates:
\begin{equation}
\frac{\mathrm{SFR (FIR)}}{M_\odot \ \mathrm{yr}^{-1}} = \frac{L_\mathrm{FIR}}{1.134\times 10^{36} \mathrm{W}}.
\label{eq:sfrfir}
\end{equation}

One has to keep in mind that the far-infrared (FIR) luminosity $L_\mathrm{FIR}$ traces the star formation rate on 100 Myr-timescales and can therefore not directly be compared to star formation rates from the often used estimators H$\alpha$ and Br$\gamma$ that trace recent star formation. Furthermore, the far-infrared luminosity $L_\mathrm{FIR}$, particularly if measured in galaxy-wide apertures, can be heavily affected by AGN emission and should therefore rather be seen as an upper limit.

\begin{figure}
\centering
\includegraphics[width=\columnwidth]{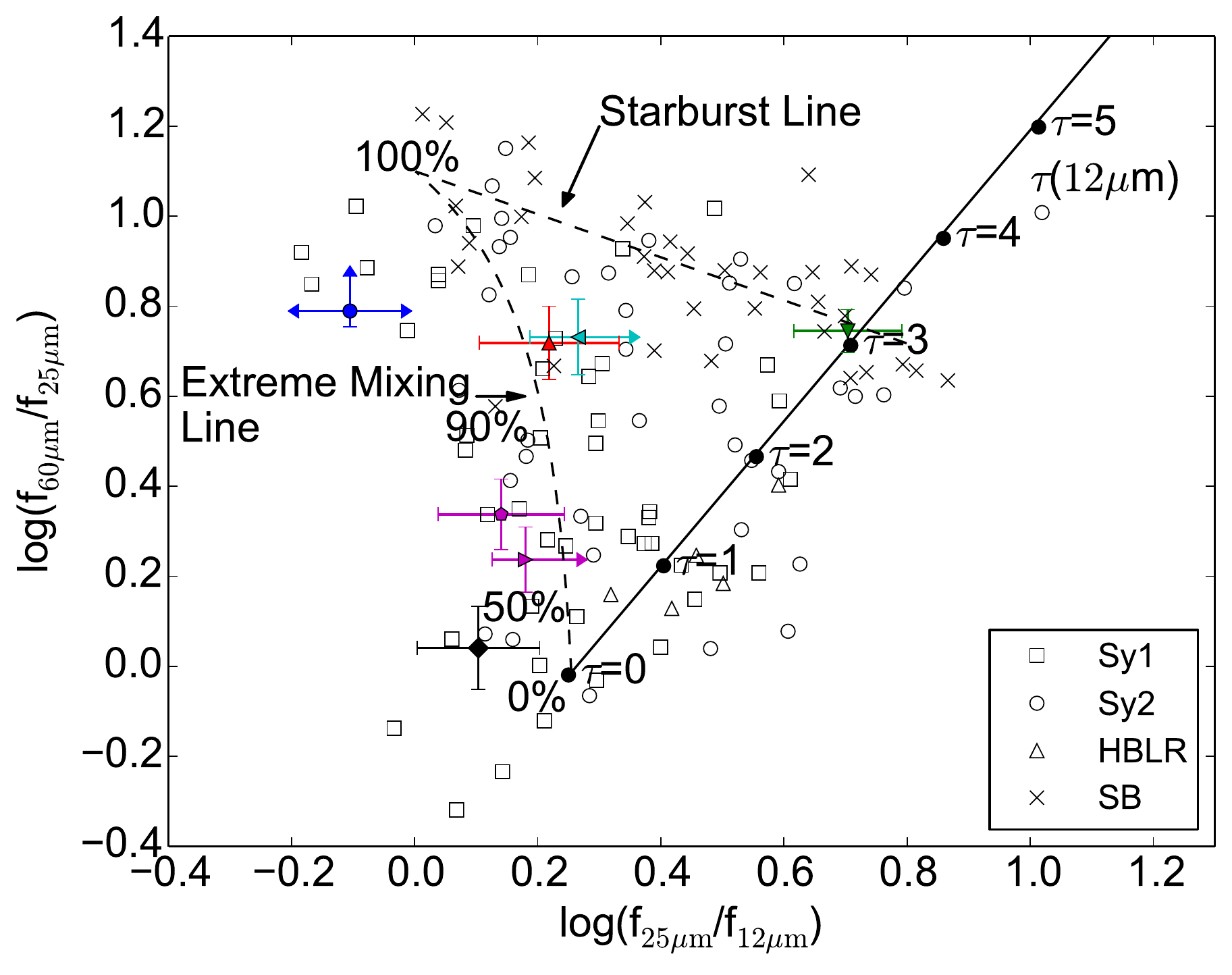}
\includegraphics[width=\columnwidth]{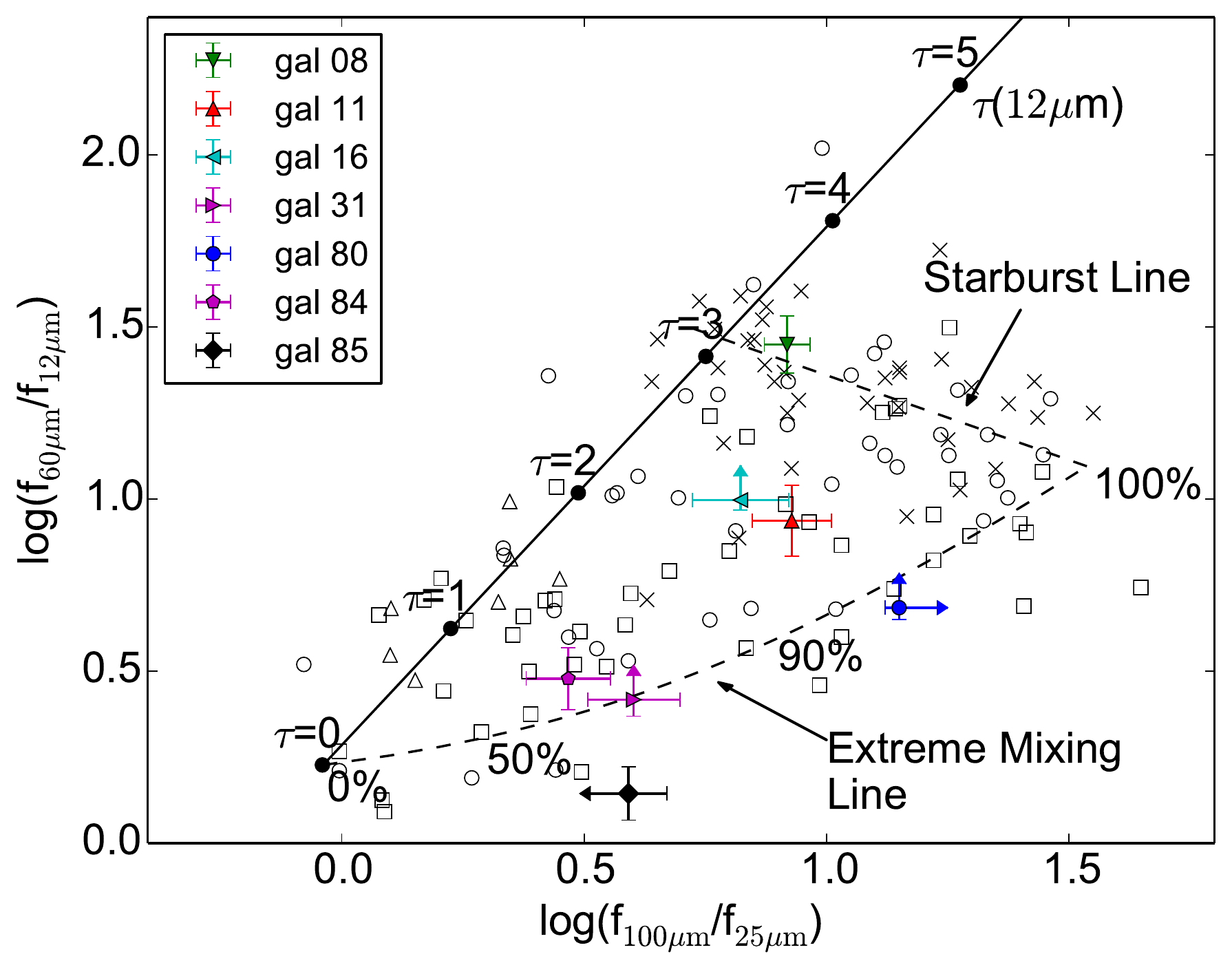}
\caption{Far-infrared colour-colour diagrams, modified from \citet{kewley_compact_2000}. \emph{Left:} $\log(f_{60\mm}/f_{25\mm})$ vs.~$\log(f_{25\mm}/f_{12\mm})$ diagram. \emph{Right:} $\log(f_{60\mm}/f_{12\mm})$ vs.~$\log(f_{100\mm}/f_{25\mm})$ diagram. In both plots, the solid line is the reddening line calculated by \citet{1998ApJ...498..570D}. The straight dashed line marks the typical area of starbursts. The ``Extreme Mixing Line'' corresponds to the mixture of a typical Seyfert-1 spectrum and a starburst component (with denoted percentage). The open symbols are Seyfert-1, Starburst, and Seyfert-2 galaxies with (HBLR) and without (Sy2) broad line components observed in polarised light, taken from the sample of \citet{1993ApJS...89....1R}. The positions of the analysed galaxies are indicated by the coloured data points.}
\label{fig:firdiag}
\end{figure}

\new{In an attempt to constrain the contribution of the AGN to the FIR luminosity, we place the galaxies in the FIR colour-colour diagrams, taken from \citet{kewley_compact_2000} (see Fig.~\ref{fig:firdiag}). In these, the typical area of starburst galaxies is indicated by a dashed line. Furthermore, a reddening line is marked by a solid line. The mixing line corresponds to the mixture of a typical unreddened Seyfert-1 spectrum with a ``warm'' starburst.
 Although the diagrams show an ambiguity between the mixing of AGN and starburst and reddening, we can make the following statements: HE0045 is completely dominated by the starburst which is fully consistent with the findings discussed notes on the individual object in the Appendix. All other galaxies have significant contributions from the AGN. For HE0103, HE0119, and HE2112, this contribution is less than 50\% while it is higher than 50\% for HE0323 and HE2211. HE2221 is completely dominated by the AGN which is consistent with the finding in the imaging study \citep{2014A&A...561A.140B} where it is the galaxy with the highest AGN fraction in the sample of 20 galaxies.}

\begin{table*}
\centering
\caption{IRAS fluxes and derived quantities: IR and FIR luminosity, FIR star formation rate.}
\label{tab:iras}
\begin{tabular}{cccccccc} \hline \hline
source & $F_{12\mm}$ & $F_{25\mm}$ & $F_{60\mm}$ & $F_{100\mm}$ & $L_\mathrm{IR}$ & $L_\mathrm{FIR}$ & SFR$_\mathrm{FIR}$ \\
 & [mJy] & [mJy] & [mJy] & [mJy] & [$10^{10}\,L_\odot$] & [$10^{10}\,L_\odot$] & [$M_\odot\,\mathrm{yr}^{-1}$] \\
 \hline

08 HE0045--2145 & $127\pm23$ & $642\pm58$ & $3570\pm220$ & $5320\pm320$ & $9.6\pm0.4$ & $5.0\pm0.2$ & 17 \\
11 HE0103--5842 & $170\pm36$ & $281\pm43$ & $1470\pm160$ & $2380\pm260$ & $7.0\pm0.5$ & $3.1\pm0.3$ & 10 \\
16 HE0119--0118 & $\leq 147$ & $271\pm49$ & $1460\pm100$ & $1800\pm250$ & $\leq 31$ & $13\pm1$ & 44 \\
31 HE0323--4204 & $\leq 111$ & $168\pm21$ & $290\pm32$ & $672\pm121$ & $\leq 17$ & $3.8\pm0.4$ & 13 \\
80 HE2112--5926 & $\leq 121$ & $\leq 95$ & $585\pm47$ & $1340\pm90$ & $\leq 5.4$ & $2.2\pm0.2$ & 7 \\
84 HE2211--3903 & $264\pm48$ & $365\pm55$ & $794\pm79$ & $1070\pm140$ & $15\pm2$ & $3.8\pm0.3$ & 13 \\
85 HE2221--0221 & $208\pm29$ & $264\pm48$ & $290\pm32$ & $\leq 1030$ & $\leq 22$ & $\leq 4.6$ & --- \\

\hline
\end{tabular}
\end{table*}

\section{Conclusions}
\label{sec:conclusions}

In this paper, we analysed near-infrared $H+K$-band longslit spectra of eleven galaxies from the \emph{low-luminosity type-1 QSO sample}. Low-resolution spectroscopy provides an insight into the gas reservoirs and possible excitation mechanisms. Our main results are the following:

\begin{enumerate}
\item All galaxies show the hydrogen recombination line Pa$\alpha$ and most galaxies also Br$\gamma$. In nine out of eleven galaxies, we detect broad components as expected for type-1 AGN. In two galaxies, only narrow components are visible. One of those, HE0045--2145, is clearly a misclassification and rather a starburst galaxy than an AGN. Therefore, it has to be removed from the LLQSO sample.

\item From the broad components of the Pa$\alpha$ emission line, we have estimated black hole masses. For those galaxies with previous black hole mass estimates, the masses agree well with our new masses. We can add four more data points in the black hole mass - bulge luminosity diagram which we discussed in \citet{2014A&A...561A.140B}, supporting the finding that LLQSOs do not follow the black hole mass - bulge luminosity relations of inactive galaxies. We discuss that the nature of this offset is essential for the understanding of the BH - host galaxy evolution.

\item From continuum fits, we derived estimates for the contributions of stars, hot dust (from obscuring torus) and power law. The stellar component is significant in all galaxies, ranging from $\sim 30\%$ (AGN dominated) to $\sim 90\%$ (dominated by stellar component). Dust temperatures are in the range of $1000-1400$ K which is typical for type-1 AGN.

\item \nnew{More than half of the galaxies show strong molecular hydrogen lines which is indicative for a large reservoir of warm, excited molecular gas.} The detection of molecular hydrogen in this study motivates deeper follow-up observations in order to study the H$_2$ excitation mechanisms in LLQSOs in more detail.

\item We analyse the gas excitation mechanisms in the galaxies using a near-infrared diagnostic diagram ($\log(\mathrm{H}_2(1-0)\mathrm{S}(1)/\mathrm{Br}\gamma)$ vs. $\log ([$\ion{Fe}{ii}$]\lambda 1.257\mm/\mathrm{Pa}\beta)$). All galaxies fall in a region populated by AGN. However, several tend towards the region of star forming galaxies, emphasising the relevance of both, star formation and AGN, in LLQSOs.
\end{enumerate}

To conclude, the here analysed spectra point at the importance of both, non-stellar continuum emission as well as stellar radiation, in the LLQSO sample. From our data, however, no clear trend can be seen which means some galaxies are clearly dominated by stellar radiation, others by non-stellar continuum emission while others are a mixture. The diversity of radiation mechanisms confirms the theory that the LLQSO sample constitutes a transition between the clearly AGN dominated QSOs at higher redshift and the probably mainly secularly evolving galaxies with - if at all - only very weak nuclear activity.

\nnew{The analysed sources are all offset from the $M_\mathrm{BH}-L_\mathrm{bulge}$ relation. At the same time, they show signs for ongoing or recent star formation, which may indicate overluminous bulges due to lower mass-to-light ratios.}

Additional multi-wavelength data and spatially resolved integral-field spectroscopy will be analysed to further investigate the interplay between AGN activity and star formation in the observed LLQSOs.

\begin{acknowledgements}
The authors thank the anonymous referee for comments that helped to improve the paper.
This work was supported by the Deutsche Forschungsgemeinschaft (DFG) via SFB 956, subproject A2. 
G.~Busch and N.~Fazeli are members of the \emph{Bonn-Cologne Graduate School for Physics and Astronomy (BCGS)}. 
S.~Smaji\'c is member of the \emph{International Max Planck Research School for Astronomy and Astrophysics Bonn/Cologne (IMPRS)}.
M. Valencia-S. received funding from the European Union Seventh Framework Programme (FP7/2007-2013) under grant agreement No. 312789.
J. Scharw\"achter acknowledges the European Research Council for the Advanced Grant Program Number 267399-Momentum. 

This publication makes use of data products from the Two Micron All Sky Survey, which is a joint project of the University of Massachusetts and the Infrared Processing and Analysis Center/California Institute of Technology, funded by the National Aeronautics and Space Administration and the National Science Foundation.

This research has made use of the NASA/IPAC Extragalactic Database (NED) which is operated by the Jet Propulsion Laboratory, California Institute of Technology, under contract with the National Aeronautics and Space Administration.
\end{acknowledgements}

\bibliographystyle{aa} 
\bibliography{llqso5}

\begin{appendix}
\section{Discussion of individual objects}
\label{sec:individual}
\new{In the following section, we discuss individual objects. Spectra of all galaxies are shown in Fig.~\ref{fig:spectra}. All spectra have been extracted from an aperture with radius corresponding to $3\times$ FWHM of the seeing.}

\subsubsection*{05 HE0036--5133}

The galaxy HE0036--5133 (also known as WPVS7) at $z=0.0288$, corresponding to $D_L=126.1\Mpc$, has been classified as Narrow-line Seyfert-1 \citep[NLS1][]{2006A&A...455..773V} and found to be X-ray transient with a broad-absorption line outflow \citep{2009ApJ...701..176L}. Fitting the hydrogen recombination line Pa$\alpha$ yields a width of the broad component of around $1300\kms$, confirming the classification as NLS1 \citep[but see also discussion about NLS1 in][]{2012nsgq.confE..17V}.

\subsubsection*{08 HE0045--2145}

HE0045--2145 (also known as ESO 540-27) is a barred spiral galaxy at redshift $z=0.0214$, corresponding to $D_L=93.2\Mpc$, i.e. one of the closest objects in our sample. An elongated, patchy structure in the direct neighbourhood indicates tidal interaction with a possibly disrupted companion. However, more detailed information about this object is missing and the galaxy itself does not show further indications for interaction. The NIR colours suggest a low AGN contribution. Prominent stellar absorption features in the analysed nuclear spectrum support the importance of the stellar contribution also in the central region. We could not find broad components in the hydrogen recombination lines that should be visible in a type-1 galaxy. 

The IRAS colours are $\displaystyle \log(f_\nu (12\mm)/ f_\nu (25\mm)) =-0.7$ and $\log(f_\nu (60\mm)/f_\nu (100\mm))=-0.17$. In the IRAS colour-colour diagrams by \cite{1986ApJ...311L..33H}, this lies in a region characteristic for starburst dominated galaxies. \nnew{Also in the FIR colour-colour diagram of \cite{kewley_compact_2000} (Fig.~\ref{fig:firdiag}), this galaxy lies on the starburst line.}

The galaxy has a \ion{H}{i} gas mass of $5.7 \times 10^9\, M_\odot$ \citep{2009A&A...507..757K} and a $\mathrm{H}_2$ gas mass of $2.9 \times 10^9\, M_\odot$ \citep{2007A&A...470..571B}. This means that stars are formed quite efficiently with a star formation efficiency (SFE) of $L_\mathrm{FIR}/M_{H_2} = 17\, L_\odot\, M_\odot^{-1}$ \citep{1994ApJ...424..627E}.

In the continuum fit (Sect.~\ref{sec:continuum}), the resulting black body temperature \nnew{of $T\sim 25\, 000\,\mathrm{K}$ is much higher than expected for a dusty torus with a temperature $T\sim 1200\,\mathrm{K}$.} This is the black body temperature of O and B stars, indicative for a recent starburst. All these arguments indicate that HE0045--2145 is rather dominated by its stellar components and ongoing star formation and that the classification as Seyfert 1 is invalid. 

\begin{figure}
\centering
\includegraphics[width=\columnwidth]{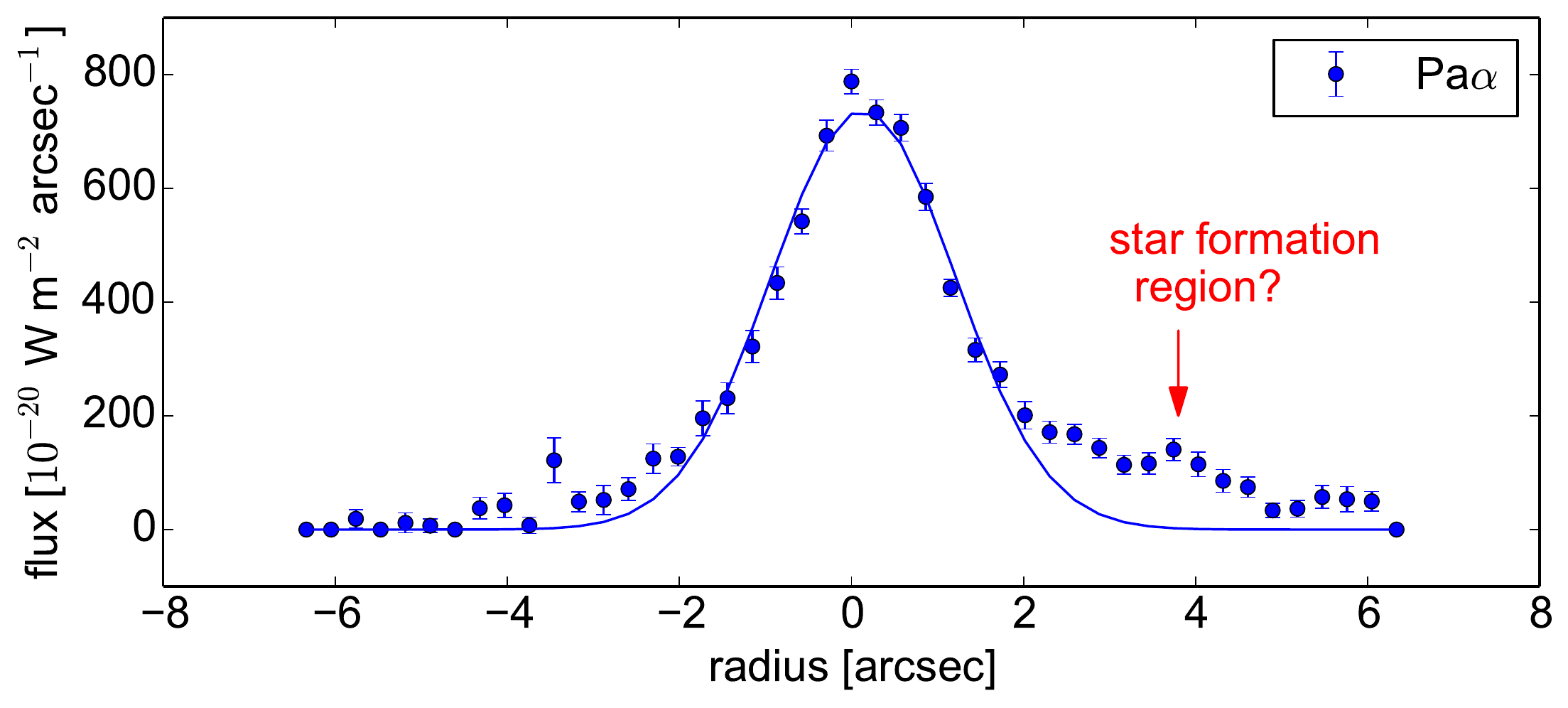}
\caption{HE0045--2145: Radial distribution of the intensity of the Pa$\alpha$ emission line. \nnew{The off-central bump at $~\sim 4\arcsec$ south-west of the nucleus (see Fig.~\ref{fig:slitplot}) could be attributed to a circumnuclear star forming region.}}
\label{fig:gal08_spat}
\end{figure}

In Fig.~\ref{fig:gal08_spat} we show the radial distribution of the Pa$\alpha$ emission which shows a bump \nnew{at a distance of $\sim 4 \arcsec$ from the center, which could be attributed to a star formation region.}

\subsubsection*{11 HE0103--5842}

HE0103--5842 (also known as ESO 113-10) has a redshift of $z=0.0257$ ($D_L=112.2\Mpc$). At first view, it looks like a barred galaxy. However, the morphological study \citep{2014A&A...561A.140B} reveals a more complex underlying structure. The galaxy has been classified as Seyfert 1.8 by \cite{2006A&A...455..773V}. No broad components could be resolved in our NIR spectra. The blue continuum and deep stellar absorption features indicate that the nuclear non-stellar component is weak in comparison to the stellar component. In the continuum fit, the stellar component has indeed a fraction of $\sim 90\%$. 

\cite{2012A&A...542A..30M} measure the Balmer decrement and find \nnew{a significant amount of reddening} in the line of sight. However, \nnew{neither our NIR Pa$\alpha$/Br$\gamma$ line ratios nor our continuum fit show} a significant contribution from extinction.

\subsubsection*{16 HE0119--0118}

HE0119--0118 (also known as Mrk1503 or II Zw 1) is a barred spiral galaxy at $z=0.0547$, corresponding to $D_L=244\Mpc$ and can be classified as luminous infrared galaxy (LIRG) based on the IRAS fluxes. The IRAS colours of $\log(f_\nu (12\mm)/f_\nu (25\mm)) \leq -0.3$ and $\log(f_\nu (60\mm)/f_\nu (100\mm)) = -0.1$ lie in the region of starburst dominated galaxies. \nnew{In the FIR colour-colour diagrams of \cite{kewley_compact_2000} (Fig.~\ref{fig:firdiag}), the galaxy consistently shows a starburst contribution of at least 50\%.} The $\mathrm{H}_2$ mass is $M_{\mathrm{H}_2}=5.4 \times 10^9 M_\odot$ \citep{2007A&A...470..571B}. This results in a star formation efficiency of $24\, L_\odot\, M_\odot^{-1}$, almost the maximum detected SFE \nnew{found in Galactic star forming regions \citep[$\leq 30\, L_\odot\, M_\odot^{-1}$;][and references therein]{1994ApJ...424..627E}.}

The broad hydrogen recombination lines ($\mathrm{FWHM}_\mathrm{BLR} \approx 3000\kms$) and the position in the diagnostic diagram (Fig.~\ref{fig:diagndiagram}) indicate \nnew{black hole accretion and that} the classification as Seyfert galaxy \citep[Sy1.5,][]{2006A&A...455..773V} is correct. However, the mentioned arguments show that star formation is a significant factor. 

In the imaging study \citep{2014A&A...561A.140B}, we found the central stellar component (bulge) to be very compact and we were not able to disentangle it from the nuclear non-stellar component. This could indicate the presence of a compact, disky pseudo-bulge. These objects are products of secular evolution and often star-forming, contrary to classical bulge components \citep[see review of][]{2004ARA&A..42..603K}. The coexistence of prominent non-stellar (e.g. broad lines) and stellar (CO absorption) features in the spectrum indicate that even in the nuclear region, star formation could be of importance, motivating more detailed investigation of this source, e.g. by integral field spectroscopy.

\nnew{The black hole mass estimates for this source} range from $\log(M_\mathrm{BH}/M_\odot)=7.1$ \citep{2012AJ....143...83X} over $\log(M_\mathrm{BH}/M_\odot)=7.58$ \citep{2011ApJ...726...59B} and $\log(M_\mathrm{BH}/M_\odot)=7.9$ (Schulze, priv. communication) to $\log(M_\mathrm{BH}/M_\odot)=7.9$ (this study), demonstrating the uncertainty of black hole mass estimates.

\subsubsection*{31 HE0323--4204}

HE0323--4204 is classified as Seyfert 1.5 by \cite{2006A&A...455..773V}. \nnew{The prominent broad emission line components of $\mathrm{Pa}\alpha$ and $\mathrm{Br}\gamma$, the significant power-law and hot dust contribution to the continuum flux,} and the forbidden [\ion{Fe}{ii}] line indicate strong nuclear activity.

The IRAS colours are $\log(f_\nu (12\mm)/f_\nu (25\mm))\leq -0.18$ and $\log(f_\nu (60\mm)/f_\nu (100\mm))=-0.36$, i.e. between the typical colours of starburst galaxies and elliptical galaxies. \nnew{In the FIR colour-colour diagrams (Fig.~\ref{fig:firdiag}), they lie on the Extreme Mixing Line with a starburst contribution of $\gtrsim 50\%$. The continuum fit yields a stellar contribution of $\sim 30\%$.} Furthermore, stellar absorption lines and the position in the diagnostic diagram (Fig.~\ref{fig:diagndiagram}) close to the location of starburst galaxies indicate an at least moderate contribution of star formation.

The galaxy at redshift $z=0.058$ ($D_L=259.3\Mpc$) is very inclined and probably a spiral galaxy. In direct neighbourhood, there is a companion galaxy that shows signs of interaction. \nnew{In some astronomical databases, the galaxy is sometimes confused with another spiral galaxy in the neighbourhood. This galaxy also falls into the slit but shows no signs for non-stellar contribution.}

\subsubsection*{80 HE2112--5926}

HE2112--5926 is part of the interacting galaxy pair ESO144-21. It has an elliptical shape, the interacting partner is a spiral galaxy. The source has redshift $z=0.0317$ ($D_L=139.1$).

The stellar contribution to the continuum emission is $\sim 80\%$. This is in agreement with the low luminosity fraction of the non-stellar point source found in the imaging study \citep{2014A&A...561A.140B}.

\subsubsection*{81 HE2128--0221}

HE2128--0221 appears as an elongated elliptical at redshift $z=0.0528$ ($D_L=235.2\Mpc$) and is the least luminous object observed in this study. This explains the \nnew{low signal-to-noise ratio}. The continuum is blue and shows clear stellar absorption lines. The hydrogen recombination line $\mathrm{Pa}\alpha$ is weak compared to the continuum emission. However, a broad component is clearly seen. The $\mathrm{Br}\gamma$ line almost vanishes in the noise and shows a strange (boxy) shape.

\subsubsection*{82 HE2129--3356}

HE2129--3356 is an elliptically shaped galaxy at redshift $z=0.0293$ ($D_L=128.3\Mpc$), classified as Seyfert 1.2 by \cite{2006A&A...455..773V}. The galaxy is surrounded by several objects it is most probably interacting with. Stellar absorption features are present and the stellar component contributes $\sim 75\%$ to the continuum emission. Nevertheless, the position in the diagnostic diagram (Fig.~\ref{fig:diagndiagram}) is that of an AGN-dominated galaxy and the presence of the coronal line [\ion{Si}{vi}] emission indicates a type-1 AGN. Also the hydrogen recombination lines are very strong and show the broadest line widths ($\sim 5000\kms$) in this study.

\subsubsection*{83 HE2204--3249}

HE2204--3249, also known as ESO 404-29, has a redshift of $z=0.0594$ ($D_L=265.9\Mpc$) and is the most distant object observed in this study. In the imaging study \citep{2014A&A...561A.140B}, the decomposition revealed a non-axisymmetrical residuum which could be interpreted as dust lane. From the line ratio Pa$\alpha$/Br$\gamma$, we get a high extinction of $A_V\approx 50$ mag.
\footnote{Assuming a case B recombination scenario, and the low-density limit with typical temperatures of $T=10\,000$ K, we expect a line ratio of about Pa$\alpha$/Br$\gamma=12.5$ \citep{2006agna.book.....O}. Furthermore, assuming a dust-screen-model $f_{obs}(\lambda)=f_{intr}(\lambda) \times 10^{-0.4 A_\lambda}$ and the mean $R_V$-dependent extinction law by \cite{1989ApJ...345..245C} $\langle A(\lambda)/A(V) \rangle = a(\lambda) + b(\lambda)/R_V$ with $R_V=3.1$, $a(\lambda)=0.574 \lambda^{-1.61}$ and $b(\lambda)=-0.527 \lambda^{-1.61}$, by using hydrogen recombination lines we can estimate the extinction 
$A_V = \frac{-2.5 \log \left( (f_{Pa\alpha}' / f_{Br\gamma}' )/(f_{Pa\alpha} / f_{Br\gamma} ) \right) }{ a(\lambda_{Pa\alpha}) + b(\lambda_{Pa\alpha})/R_V - a(\lambda_{Br\gamma}) + b(\lambda_{Br\gamma})/R_V}$
with observed fluxes $f_{Pa\alpha}',f_{Br\gamma}'$ and the expected line ratio $f_{Pa\alpha} / f_{Br\gamma} \approx 12.5$.}
The continuum fit shows no sign of extinction (but see comment on reliability of this fit in Sect.~\ref{sec:continuum}).

Several objects \nnew{located in direct neighbourhood are probably in a tidal interaction with the galaxy. The spectrum shows a very blue continuum shape and} deep stellar absorption lines are seen. From the continuum fit we get a stellar contribution to the continuum of about $90\%$. Additionally, the hydrogen recombination lines are rather weak compared to the continuum flux. Nevertheless, we detect broad components ($\mathrm{FWHM}_\mathrm{BLR}\approx4000 \kms$), consistent with the classification as Seyfert 1.2 in \cite{2006A&A...455..773V}.

\subsubsection*{84 HE2211--3903}

HE2211--3903 (also known as ESO 344-16) is a barred spiral galaxy, classified as Seyfert 1.5 galaxy by \cite{2006A&A...455..773V}. The redshift is $z=0.0398$, corresponding to $D_L=175.6\Mpc$. In the previous imaging study \citep{2014A&A...561A.140B}, we reveal an additional spiral arm and an inner ring that can also be found in emission line maps by \cite{2011AJ....142...43S}. 

Confirming the results from \citet{2006A&A...452..827F}, the spectrum is very red, due to black body radiation coming from hot dust at a temperature of $\sim 1100$ K which makes up about $60\%$ of the continuum radiation. Probably due to this high contribution of hot dust, the spectrum does not show any other features than the hydrogen recombination lines $\mathrm{Pa}\alpha$ and Br$\gamma$. 

\citet{2011AJ....142...43S} report that based on unresolved narrow-line ratios ($\log([$\ion{N}{ii}$]/\mathrm{H}\alpha)=-0.2$, $\log([$\ion{O}{iii}$]/\mathrm{H}\beta)=0.2$ and $\log([$\ion{S}{ii}$]/\mathrm{H}\alpha)=-0.5$), the galaxy would be classified as ``composite''. However, through spatially resolved spectroscopy, they find an extended narrow-line region on scales up to $8\,\mathrm{kpc}$.

Based on IRAS fluxes, the galaxy is a LIRG. The $\mathrm{H}_2$ mass is $M_{\mathrm{H}_2}=9.3 \times 10^9 M_\odot$ \citep{2007A&A...470..571B}, resulting in a SFE of $4\, M_\odot\, L_\odot^{-1}$. We see that the IRAS data suggest ongoing star formation activity. \nnew{Since there is no evidence for any strong interaction, HE2211--3903 is likely to be an example for a secularly evolving galaxy.}

\subsubsection*{85 HE2221--0221}

HE2221--0221 is also known as 3C 445 and classified as a ``double double'' radio galaxy \citep{2000MNRAS.315..371S}. It has a redshift of $z=0.057$ ($D_L=254.7\Mpc$). In near-infrared images, it appears as round elliptical with very bright unresolved nucleus. Our imaging study \citep{2014A&A...561A.140B} shows that the host is almost over-shone by the nucleus. The $\mathrm{Pa}\alpha$ line is very broad and strong. However, the $\mathrm{Br}\gamma$ line is rather weak. This could be induced by reddening. Indeed, the shape and the NIR colours are extremely red and the continuum fit yields an extinction of $A_V=1.6$ mag. Nevertheless, \nnew{the spectrum shows some stellar features,} and the stellar contribution is around $\sim 30\%$. The presence of strong [\ion{Si}{vi}] emission is indicative of \nnew{nuclear activity}.

\begin{figure*}
\centering
\includegraphics[width=\linewidth]{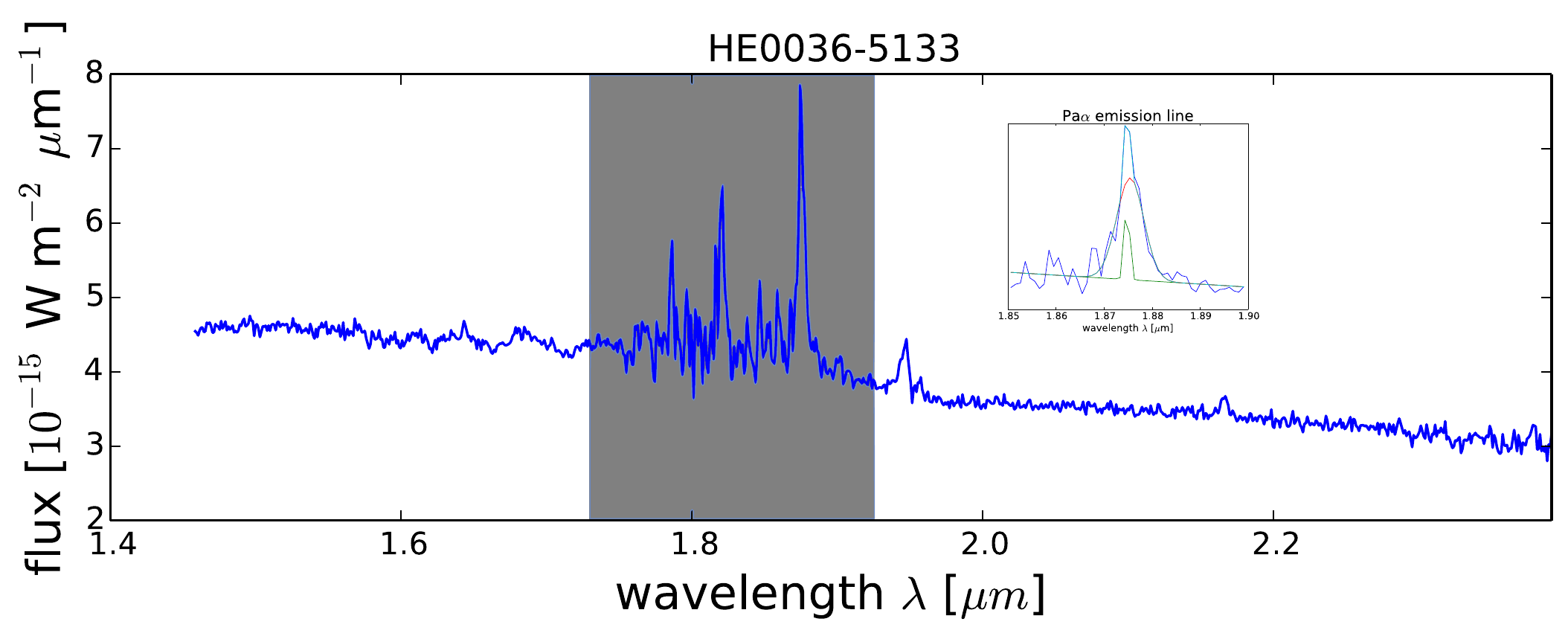}
\includegraphics[width=\linewidth]{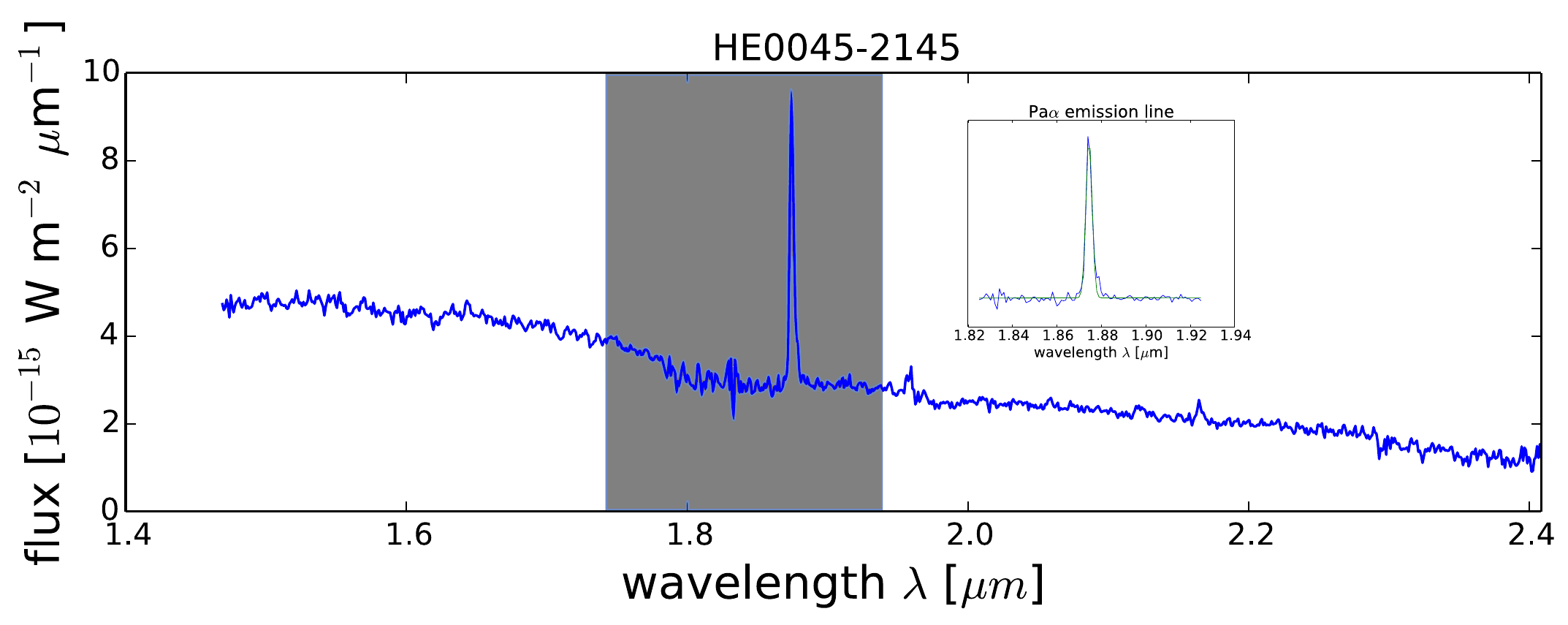}
\includegraphics[width=\linewidth]{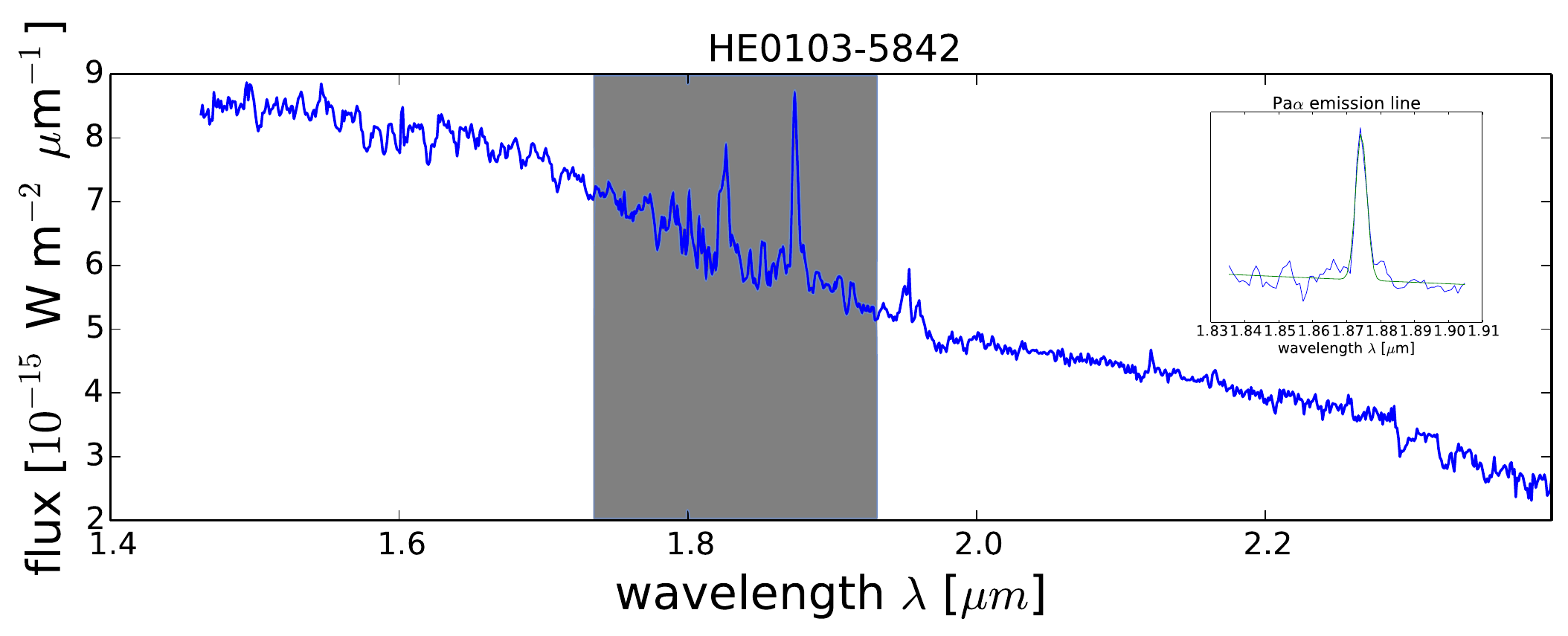}
\caption{Spectra of all analysed galaxies. Spectra have been extracted from an aperture with radius corresponding to $3\times$ FWHM. The region between $H$- and $K$-band with low transmission has been marked in grey.}
\label{fig:spectra}
\end{figure*}

\begin{figure*}
\ContinuedFloat
\centering
\includegraphics[width=\linewidth]{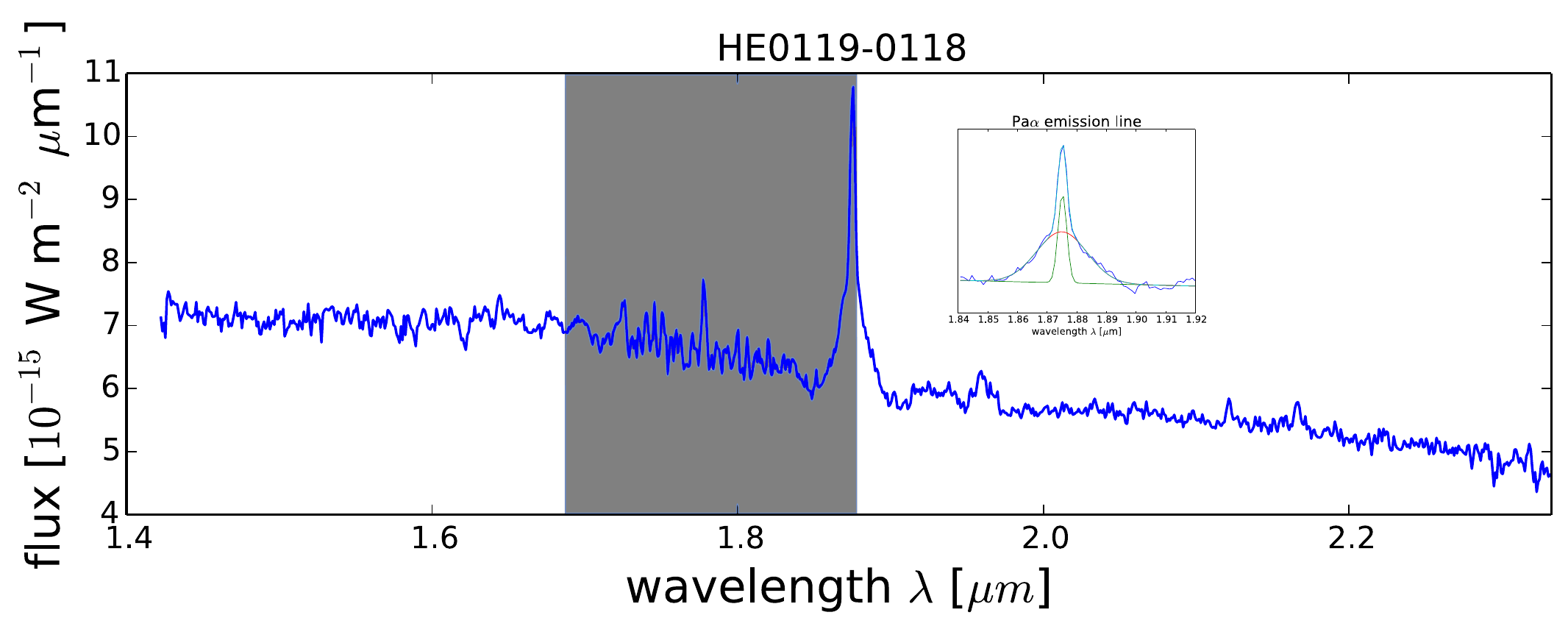}
\includegraphics[width=\linewidth]{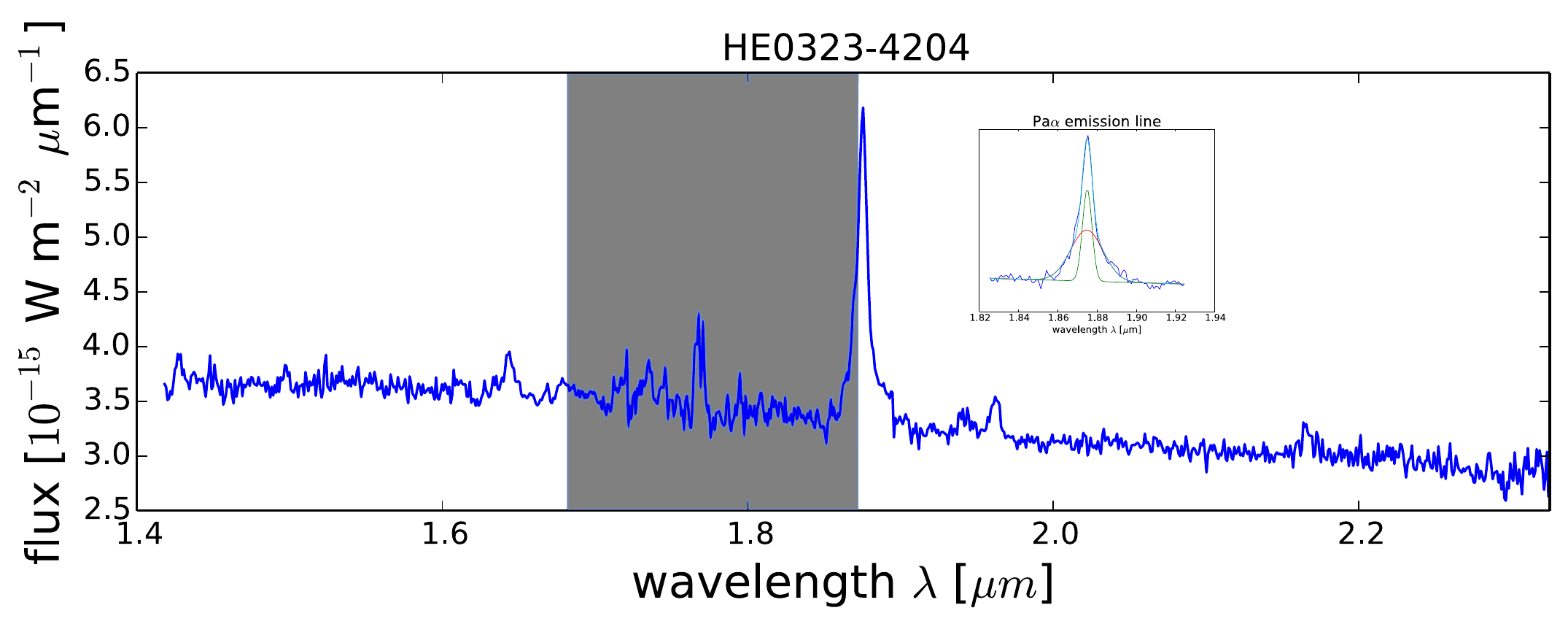}
\includegraphics[width=\linewidth]{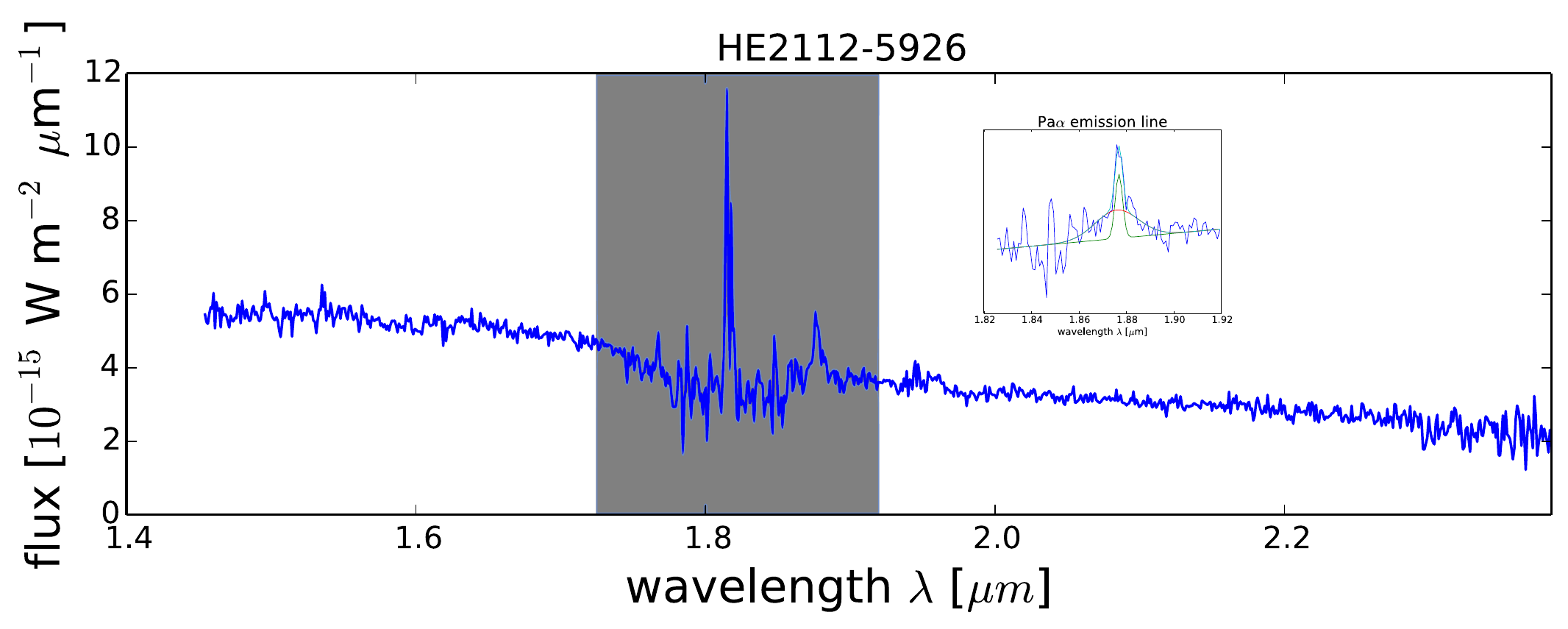}
\caption{continued.}
\end{figure*}

\begin{figure*}
\ContinuedFloat
\centering
\includegraphics[width=\linewidth]{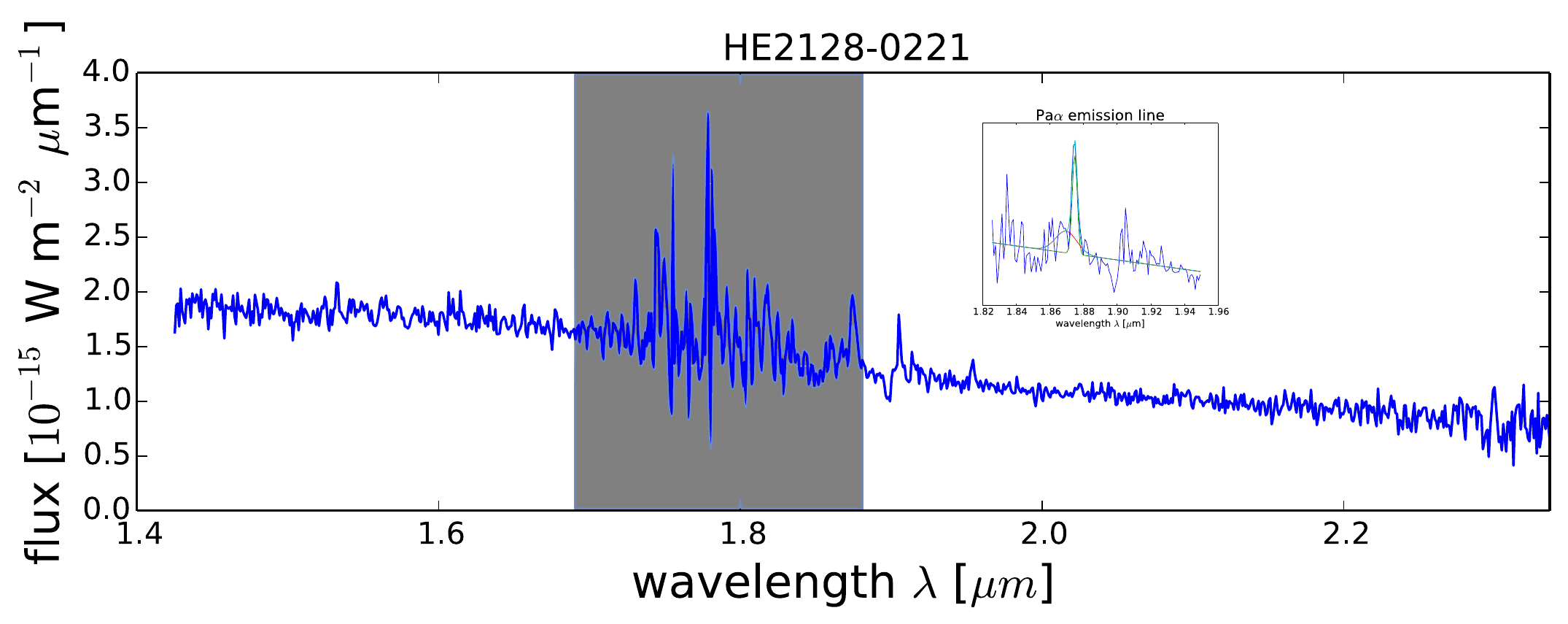}
\includegraphics[width=\linewidth]{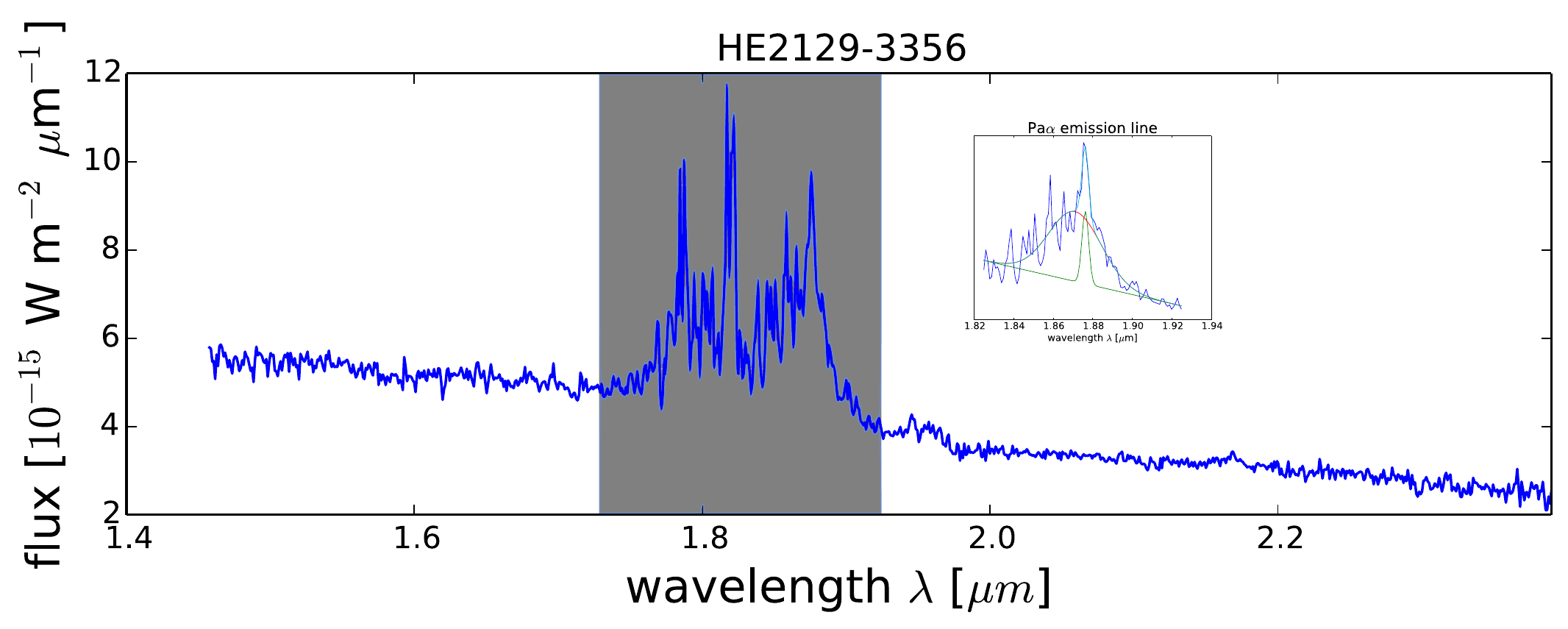}
\includegraphics[width=\linewidth]{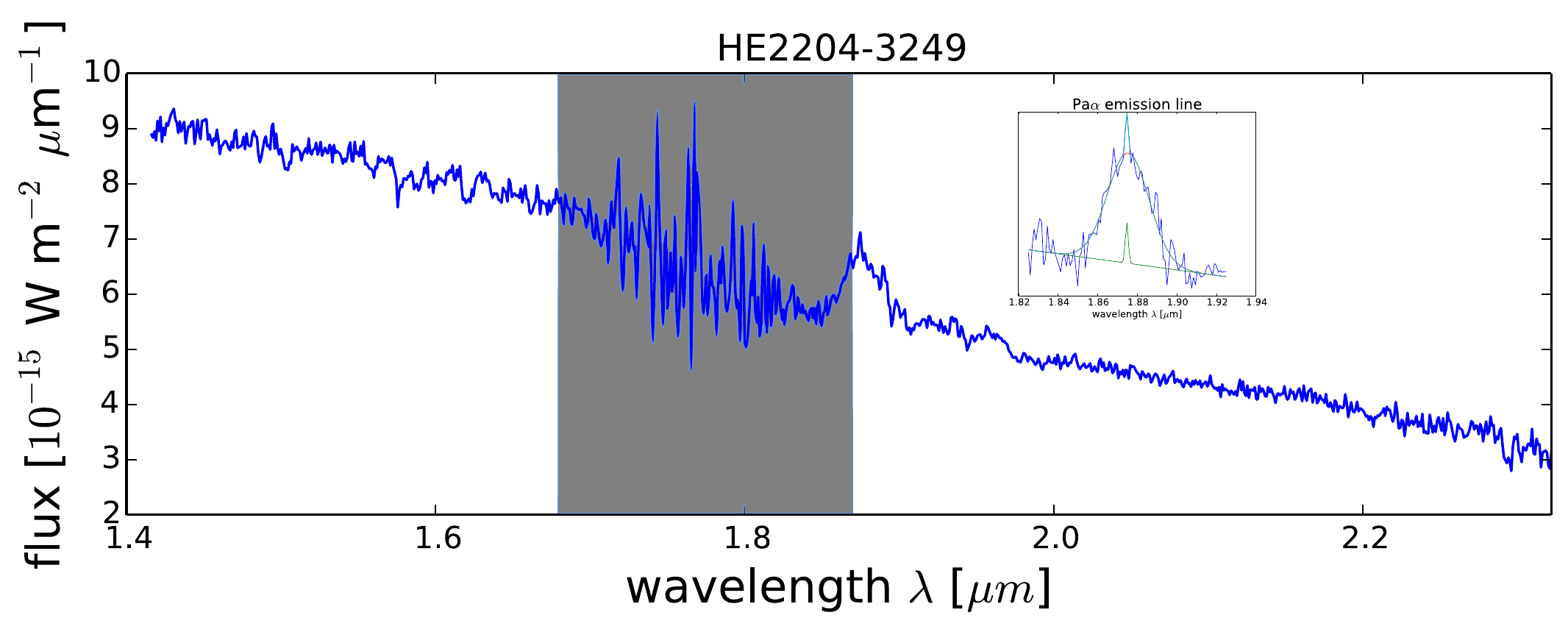}
\caption{continued.}
\end{figure*}

\begin{figure*}
\ContinuedFloat
\centering
\includegraphics[width=\linewidth]{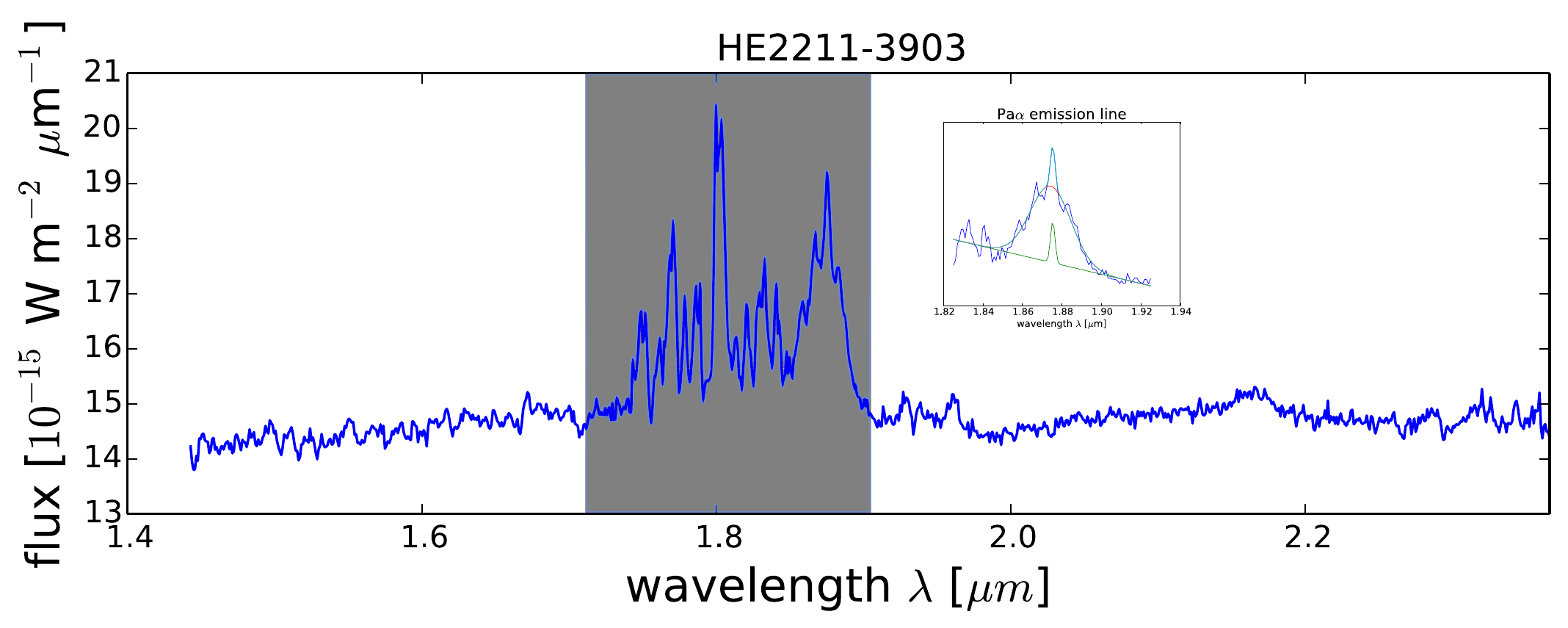}
\includegraphics[width=\linewidth]{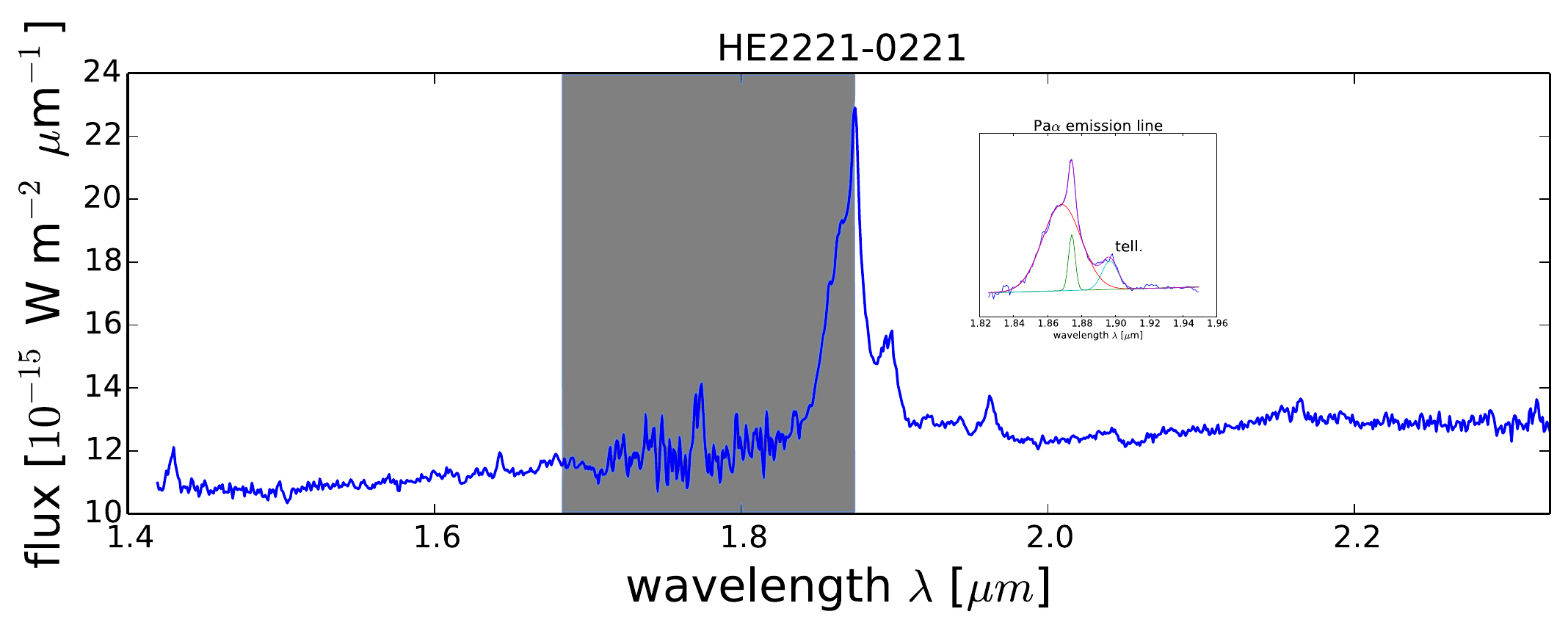}
\caption{continued.}
\end{figure*}

\end{appendix}

\end{document}